\documentclass[10pt,journal,compsoc]{IEEEtran}
\usepackage{graphicx}
\usepackage{booktabs}
\usepackage{subcaption}
\usepackage{epstopdf}
\usepackage{diagbox}
\usepackage{threeparttable}
\usepackage{makecell,multirow,diagbox}

\usepackage[linesnumbered,ruled,vlined]{algorithm2e}
\usepackage{algpseudocode}
\usepackage{amssymb}
\usepackage{amsmath}
\usepackage{amsthm}
\usepackage{cite}
\usepackage{hyperref}
\usepackage{bm}
\usepackage{mathtools}
\usepackage{pgfmath}
\usepackage{dsfont}
\usepackage{xcolor}
\usepackage{color}

\usepackage{flushend}

\def\ourlayer{MonoConv}
\def\ourmethod{MonoCNN}

\def\etc{etc.\@\xspace}
\def\ie{i.e.,\@\xspace}

\ifCLASSINFOpdf

\else

\fi

\begin{document}
%

\title{Towards Transmission-Friendly and Robust CNN Models over Cloud and Device}

\author{Chuntao~Ding,
        ~Zhichao~Lu,
        ~Felix~Juefei-Xu,~\IEEEmembership{Member,~IEEE,}
        ~Vishnu~Naresh~Boddeti,~\IEEEmembership{Member,~IEEE,}
        ~Yidong~Li,~\IEEEmembership{Senior Member,~IEEE,}
        ~Jiannong~Cao,~\IEEEmembership{Fellow,~IEEE}\\
\IEEEcompsocitemizethanks{\IEEEcompsocthanksitem Chuntao Ding and Yidong Li are with the School of Computer and Information Technology, Beijing Jiaotong University, Beijing, China.
E-mail: \{chtding, ydli\}@bjtu.edu.cn.
\IEEEcompsocthanksitem Zhichao Lu is with the School of Software Engineering, 
                Sun Yat-sen University, Zhuhai 519082, China.
                E-mail: luzhichaocn@gmail.com.\\
(Corresponding author: Zhichao Lu)
\IEEEcompsocthanksitem F. Juefei-Xu is with Alibaba Group, USA. E-mail: juefei.xu@gmail.com.
\IEEEcompsocthanksitem Vishnu N. Boddeti is with the Department of Computer Science and Engineering,
                Michigan State University, East Lansing, MI, 48824, USA.
                E-mail: vishnu@msu.edu.

\IEEEcompsocthanksitem Jiannong Cao is with the Department of Computing, The Hong Kong Polytechnic University, Hong Kong, China.
E-mail: {csjcao@comp.polyu.edu.hk}.
}
}

\markboth{IEEE Transactions on Mobile Computing}%
{Shell \MakeLowercase{\textit{et al.}}: Bare Demo of IEEEtran.cls for IEEE Transactions on Magnetics Journals}


\IEEEtitleabstractindextext{%
\begin{abstract}
Deploying deep convolutional neural network (CNN) models on ubiquitous Internet of Things (IoT) devices has attracted much attention from industry and academia since it greatly facilitates our lives by providing various rapid-response services. Due to the limited resources of IoT devices, cloud-assisted training of CNN models has become the mainstream. However, most existing related works suffer from {\emph{a large amount of model parameter transmission and weak model robustness}}. To this end, this paper proposes a cloud-assisted CNN training framework with low model parameter transmission and strong model robustness. In the proposed framework, we first introduce \ourmethod{}, which contains only a few learnable filters, and other filters are nonlearnable. These nonlearnable filter parameters are generated according to certain rules, i.e., the filter generation function (FGF), and can be saved and reproduced by a few random seeds. Thus, the cloud server only needs to send these learnable filters and a few seeds to the IoT device. Compared to transmitting all model parameters, sending several learnable filter parameters and seeds can significantly reduce parameter transmission. Then, we investigate multiple FGFs and enable the IoT device to use the FGF to generate multiple filters and combine them into \ourmethod{}. Thus, \ourmethod{} is affected not only by the training data but also by the FGF. The rules of the FGF play a role in regularizing the \ourmethod{}, thereby improving its robustness. Experimental results show that compared to state-of-the-art methods, our proposed framework can reduce a large amount of model parameter transfer between the cloud server and the IoT device while improving the performance by approximately 2.2\% when dealing with corrupted data. The code is available at \url{https://github.com/zhichao-lu/mono-cnn-pytorch}.
\end{abstract}

\begin{IEEEkeywords}
Internet of Things, cloud computing, cloud-assisted, CNNs.
\end{IEEEkeywords}}

\maketitle

\IEEEpeerreviewmaketitle

\section{Introduction}
\noindent
{\bf{Background \& Motivation.}}
With the advent of the Internet of Everything era hundreds of millions of Internet of Things (IoT) devices will be connected to the network. With the excellent performance of the deep convolutional neural networks (CNNs) in computer vision~\cite{Simonyan@Very, He@Deep}, speech~\cite{Alexander@Exploiting}, natural language processing~\cite{Nal@A, Shervin@Deep}, deploying CNNs on IoT devices can provide various convenient services~\cite{Hu@Network, Wu@Collaborate, Lu@Augur, Amir@JointDNN, Liang@A, Qian@Deep}. Limited by the insufficient resources of IoT devices, the method to successfully benefit from the excellent performance of CNNs is to seek well-resourced cloud servers to assist in training CNN models.

\begin {figure}[t]
\centering
\includegraphics[width=0.9\linewidth]{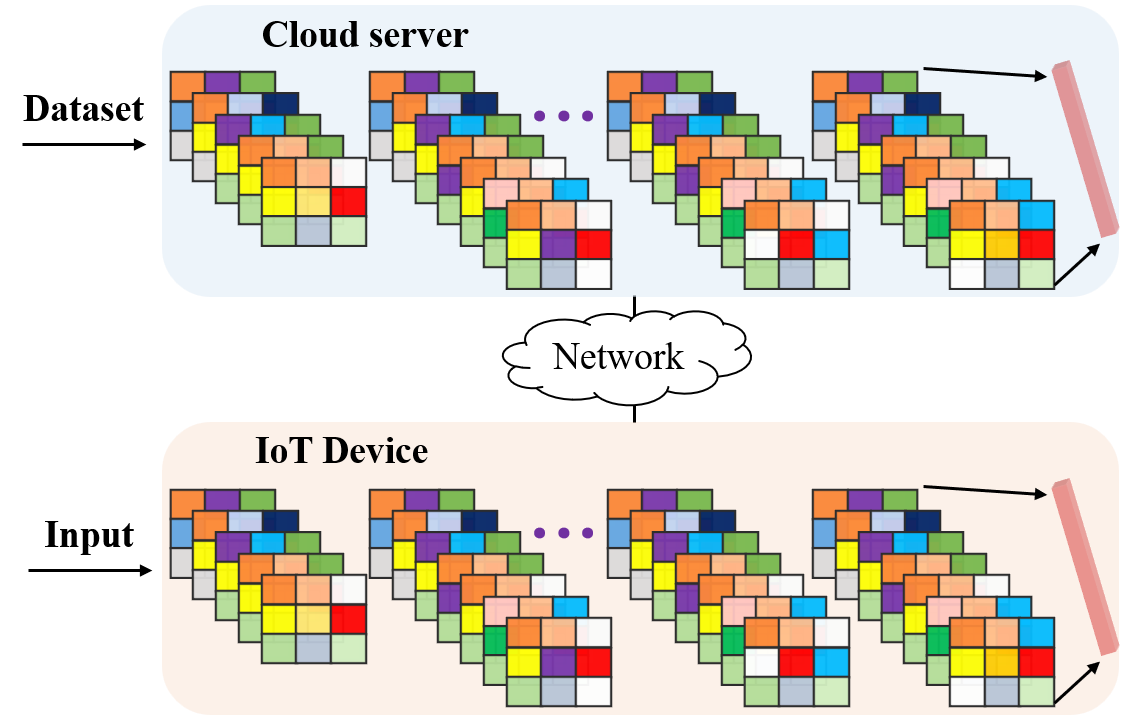}
\caption {System architecture for cloud-assisted training of CNN models.}
\label{fig:problem}
\vspace{-0.5cm}
\end{figure}

Fig.~\ref{fig:problem} shows the process of cloud-assisted CNN model training. The system architecture consists of two components: IoT devices and cloud servers. The deep CNN model is trained in the cloud server and then sent to the IoT device to provide users with services. When the subsequent CNN model is updated, the cloud server will periodically deliver the updated model to the IoT device. To combine ubiquitous IoT devices with high-performance CNN models and provide users with high-quality services this paper will study the cloud-assisted training of CNN models.

\vspace{3pt}
\noindent\textbf{Challenges.}
Implementing cloud-assisted training of a CNN model system is a nontrivial task that faces the following two key challenges: The first key challenge is to \emph{reduce the model parameters} sent by the cloud server to the IoT devices. The cloud server usually assists hundreds of millions of IoT devices in deploying and updating CNN models. In addition, the number of parameters of the deep CNN model is high. During model training or subsequent model updating, frequent and large numbers of model parameter exchanges will place considerable pressure on the network bandwidth. Therefore, reducing the amount of model parameter transmission is a prerequisite for the smooth progress of cloud-assisted training of CNN models in the era of the Internet of Everything.
    
The second key challenge arises from \emph{enhancing the robustness} of the CNN model on the IoT device. Due to the universality of the distribution of IoT devices, the input data for the execution of tasks are prone to degradation of CNN model performance due to environmental influence or man made malicious attacks. For example, the image data obtained on a rainy or a snowy day or the image is slightly enlarged, or some pixels are removed. Ensuring the robustness of the CNN model is a practical problem. The robustness of a model in this paper refers its generalization performance against corrupted data. Therefore, ensuring the robustness of the CNN models deployed on IoT devices is the key to its deployment.

\vspace{3pt}
\noindent\textbf{Our solutions.}
To address the first challenge, we propose \ourmethod{}. In \ourmethod{}, we only learn a {\emph{single filter in each layer}}, referred to as the seed filter, and generate the other parameters of the layer through a seed filter and filter generation function (FGF). The parameters of FGF are randomly generated and fixed, which allows them to be reproducible with a few random seeds. Therefore, the cloud server only needs to send these seed filters and random seeds to the IoT device, and the trained \ourmethod{} model can be reproduced on the IoT device. Compared with sending all the model parameters, sending these seed filters and seeds can significantly reduce the number of parameters transmitted from the cloud server to the IoT device.

To address the second challenge, we propose that the parameters of the \ourmethod{} do not completely depend on the training data. In the \ourmethod{}, only the parameters of the seed filter are obtained through training, and the other parameters are obtained through FGF. This makes \ourmethod{} affected not only by the training data but also by the rules of the FGF. As a result, our \ourmethod{} naturally avoids overfitting through FGF regularization so that it has better generalization when inputting corrupted data. We also investigate five FGFs and find that the monomial function significantly outperforms the others.

In summary, our main contributions are as follows:
\begin{itemize}
\item To the best of our knowledge, this is the first work that seeks to reduce model parameter transmission when training CNN models in a cloud-assisted way. Our key idea is to issue only a small number of seed filters and seeds and improve model robustness by incorporating filter generation function rules.

\item We perform a theoretical analysis of the MonoConv layer, showing that it can approximate the standard convolutional layer well.

\item The experimental results show that the proposed framework reduces a large amount of model parameter transfer between the cloud server and the IoT device and improves the mean accuracy by approximately 2.2\% when dealing with corrupted data.
\end{itemize}

The rest of the paper is organized as follows: Section~\ref{ref:2-related} reviews related work. Section~\ref{ref:3-approach} describes the proposed framework. Section~\ref{ref:4-experiment} presents our evaluation results. Finally, we conclude this paper in Section~\ref{ref:5-conclusion}.

\section{Related Work} \label{ref:2-related} 
Combining cloud servers, Internet of Things (IoT) devices, and deep neural network models to provide users with high-quality services has become mainstream. We group existing work into three categories (cloud-only, device-only, and cloud-device collaboration) based on where the neural network model training and inference are performed.

\begin {figure*}[t]
\centering
\includegraphics[width=0.8\linewidth,clip]{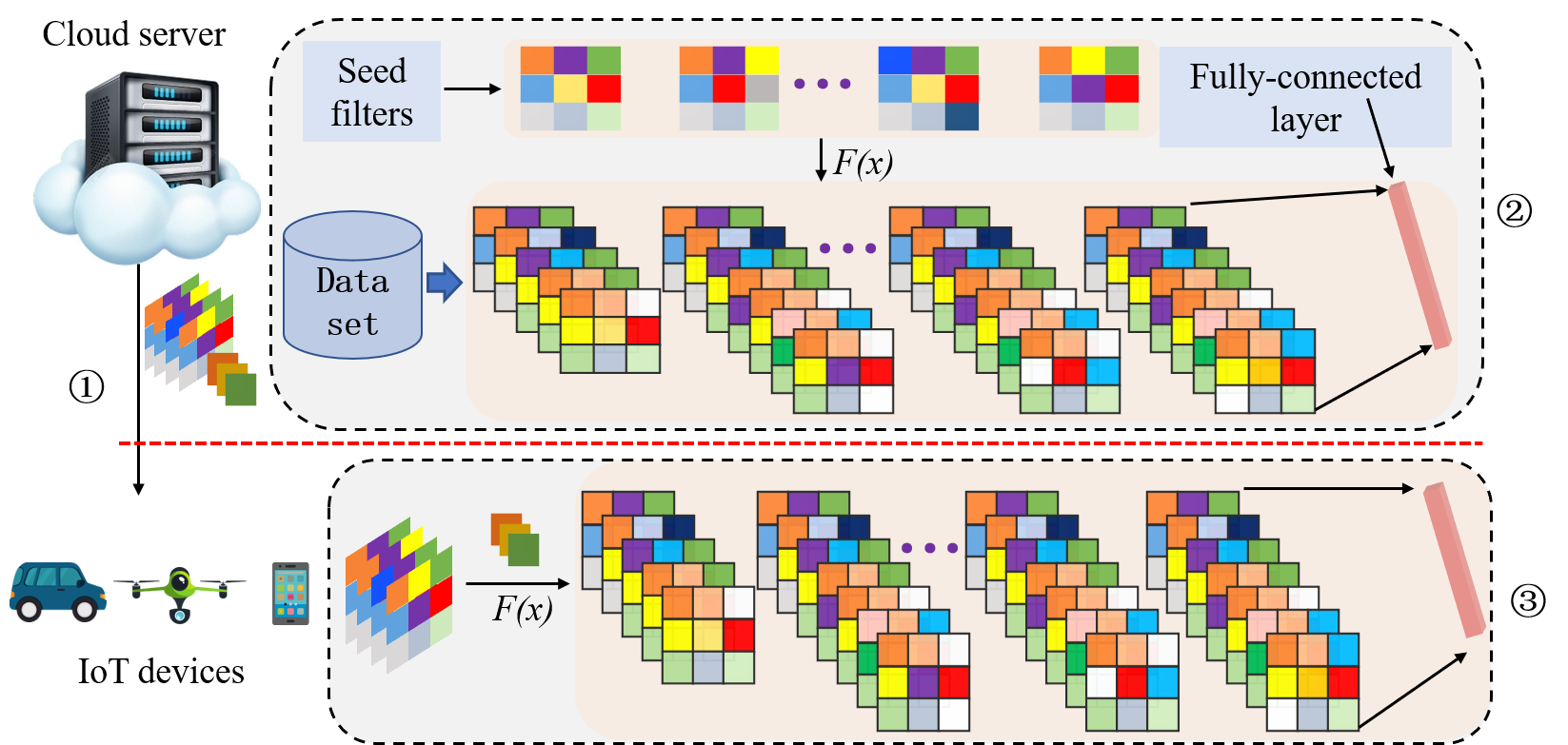} 
\caption {Overview of the proposed framework.
In the proposed framework, we first design the cloud-assisted method. Then, we design a seed filter-based CNN (i.e., \ourmethod{}). Each layer of \ourmethod{} only needs to learn the weights of one filter (i.e., seed filter), which makes the cloud server only need to send the seed filters and a small number of seeds to the IoT device. Finally, the IoT device generates the \ourmethod{} trained by the cloud server according to the seed filters, seeds and filter generation function.}
\label{fig:framework}
\vspace{-0.5cm}
\end{figure*}

\vspace{3pt}
\noindent{\bf{Cloud-only:}}
The key desiderata of generating a high-performance neural network model are sufficient computing resources and sufficient training data. The configuration of the cloud server perfectly matches these desiderata, which introduced research on running neural network models in the cloud server~\cite{Jiang@Chameleon, Liu@ANew, Georganas@Anatomy, Hazelwood@Applied}. Among them, Jiang~\emph{et al.}~\cite{Jiang@Chameleon} proposed a video analysis controller based on cloud and deep neural networks. Liu~\emph{et al.}.~\cite{Liu@ANew} proposed a deep learning-based food recognition system that runs a deep neural network model in the cloud, and the device obtains recognition services by uploading the collected data to the cloud. To reduce the quantity of data uploaded to the cloud, they also incorporate edge computing to process data on edge servers~\cite{Abbas@Mobile, Mach@Mobile, Liu@Online}. However, cloud execution is highly dependent on network conditions. When network conditions are unstable or disconnected, cloud-based deep neural network models become degraded or unavailable.

\vspace{3pt}
\noindent{\bf{Device-only:}}
With the enhancement of computing and storage capabilities of IoT devices, it is possible to train neural network models directly on IoT devices, which has also led to the birth of many excellent lightweight models, such as MobileNets~\cite{Howard@MobileNets, Sandler@MobileNetV2, Howard@Searching}, resource-aware models~\cite{Fang@NestDNN}, inference efficiency~\cite{Fang@FlexDNN, Teerapittayanon@BranchyNet} and others~\cite{Mathur@DeepEye}. For example, Howard~\emph{et al.}~\cite{Howard@MobileNets} used depthwise separable convolution instead of standard convolution to reduce the number of parameters in the network model. Fang~\emph{et al.}~\cite{Fang@NestDNN} deployed many models on end-devices and nested these models together to provide users with multiple model choices while saving storage and switching overhead. Teerapittayanon~\emph{et al.}~\cite{Teerapittayanon@BranchyNet} and Fang~\emph{et al.}~\cite{Fang@FlexDNN} introduced an early exit and multi-branch network to improve the efficiency of inference. The above methods explore how to modify the neural network model to adapt to the IoT device or better training or inference. Hence they are complementary to our proposed approach.

\vspace{3pt}
\noindent{\bf{Cloud-device collaboration:}}
Research on cloud-device collaborative training and inference has received high levels of attention with a large number of excellent approaches~\cite{Zhang@Collaborative, Stefanos@SPINN, Teerapittayanon@Distributed, Ren@Distributed, Kang@Neurosurgeon, Hu@Dynamic, Li@JALAD, Han@An} have been proposed. For example, Zhang~\emph{et al.}~\cite{Zhang@Collaborative} train the neural network model through cloud-edge collaboration and prune the deep neural network in the cloud to minimize the number of model transmission parameters while retaining the original model performance to the greatest extent. Stefanos~\emph{et al.}~\cite{Stefanos@SPINN} proposed a progressive inference method for collaborative device and cloud computing and used compression~\cite{Han@Deep} and quantization~\cite{Hubara@Quantized} to reduce the amount of parameter exchange between the device and the cloud. Kang~\emph{et al.}~\cite{Kang@Neurosurgeon} divided the CNN into a head that runs on the device and a tail that runs on the cloud and decided the split point according to the load of the device and the cloud and network conditions. Akin, Li~\emph{et al.}~\cite{Li@JALAD} proposed a joint accuracy and latency-aware execution framework, which explores the splitting points of neural network models so that one part runs on edge devices and the other part runs in the cloud, achieving fewer parameter exchanges. The above methods have made great contributions to reducing model parameter exchange. However, some methods reduce the exchange of model parameters by finding the best splitting point. Since the optimal splitting points of different models are different, it is time-consuming and labor-intensive to search for suitable splitting points. In addition, reducing the transmission of model parameters through compression and quantization results in a loss in model performance.

In contrast, our proposed approach only requires the cloud server to send one seed filter and one seed for each layer in \ourmethod{} to the IoT device, which solves the overload of network transmission bandwidth caused by excessive model parameter transmission. In addition, we also address the challenge of ensuring model robustness, which is ignored by the above approaches, by using rules to regularize the generation of model parameters.

\section{Design of the proposed approach} \label{ref:3-approach} 

\subsection{Overview}
Fig.~\ref{fig:framework} illustrates the architecture of the proposed  framework. In our framework, we first train \ourmethod{} in the cloud server. Instead of learning all the parameters of \ourmethod{}, we only learn a single filter in each layer, referred to as the seed filter. Other parameters of each layer are generated by its corresponding seed filter and filter generation function (FGF) and can be reproduced through a random seed. Therefore, the cloud server only sends a small number of seed filters and seeds to the IoT device. After obtaining seed filters and seeds, the IoT device uses them to generate multiple novel filters and combine them into the \ourmethod{} trained on the cloud server. We describe the proposed framework in detail as follows:  \textcircled{1}: the cloud-assisted method in Section~\ref{ref:cloud-device collaboration}, \textcircled{2}: seed filter learning in Section~\ref{ref:seed filter learning}, and \textcircled{3}: the filter generation function in Section~\ref{ref:filter generation function}.

\subsection{Design of cloud-assisted method} \label{ref:cloud-device collaboration} 
Our goals are threefold i) leverage the complete resources of cloud servers, ii) limit the demand for resources on IoT devices, and iii) minimize the amount of model parameter transmission. To facilitate this goal, we propose training \ourmethod{} on the cloud server first and then sending the trained model to the IoT device for deployment. In general, one cloud server corresponds to millions of IoT devices. We train \ourmethod{} on the cloud server and then send the trained \ourmethod{} to IoT devices, which facilitates updating and maintaining our model on the IoT device.

\subsection{Design of seed filter learning} \label{ref:seed filter learning}
Existing high-performance CNN models have a large number of parameters. For example, VGG19~\cite{Simonyan@Very} has 144 million parameters. In addition, one cloud server corresponds to millions of IoT devices. Thus a large number of model parameters still need to be transmitted to IoT devices. To this end, we start with an analysis of the CNN model parameters. As~\cite{han2020ghostnet} shows, the standard CNN model contains many redundant parameters~\cite{He@Deep}. To reduce the number of learnable parameters in the CNN model,~\cite{han2020ghostnet} first generates several filters and then generates some novel filters through inexpensive operations. Juefei et al.~\cite{juefei2017local} decomposed a standard convolutional layer into two modules, a nonlearnable layer, and a $1\times 1$ convolutional layer. Introducing the nonlearnable layer may represent a breakthrough in reducing the number of model parameters sent by the cloud server to the IoT devices. This is because the nonlearnable parameters are randomly initialized and can simply be saved and reproduced from a random seed. The nonlearnable parameters in this paper refer to the parameters in the CNN model that remain unchanged during training and inference and remain unchanged during model training on the cloud server.

The above analysis inspires us to specify that the parameters of only one filter (called the seed filter) in each layer of the CNN model should be learnable while the parameters of all other filters are nonlearnable and are generated per certain rules based on the seed filter. In this paper, we refer to this CNN as \ourmethod{}. Formally, in any given layer, given the seed $\bm{w}_i$ for that layer, we can generate many new filters. The filters are generated via certain specified rules, e.g., a nonlinear transformation $\bm{v} = f(\bm{w}_i)$, where $f(w^j_i)=\text{sign}(w_i^j)|w_i^j|^\beta$ is a monomial that operates on each element of $\bm{w}_i$ and $\beta > 0$ is the exponent. The convolutional outputs are computed as follows (we consider 1-D signals for simplicity):
{
\begin{align}
    \bm{y} = \sum_{j=1}^C f(\bm{w}_i^j) \ast \bm{x}^j 
\end{align}
}
where $\bm{x}^j$ is the $j^{\mathrm{th}}$ channel of the input image and $\bm{w}_i^j$ is the $j^{\mathrm{th}}$ channel of the $i^{\mathrm{th}}$ filter. During the forward pass, weights are generated from the seed filter and are then convolved with the inputs, \ie
{
\begin{align}
    z[i] &= f(w[i]) = \mathrm{sign}(w[i])|w[i]|^{\beta}  \\
    v[i] &= \frac{z[i]-\frac{1}{n}\sum_i z[i]}{\left(\sum_i\left(z[i]-\frac{1}{n}\sum_i z[i]\right)^2\right)^{\frac{1}{2}}} \label{eq:weight_normalization_start}
\end{align}
}where we normalize the response maps to prevent the responses from vanishing or exploding and $\bm v$ is the normalized response map.

Therefore, for a layer in \ourmethod{}, by specifying a seed filter along with certain rules (e.g., monomial functions), we can generate or augment as many filters as needed. For example, assume that we need $m$ filters in total for one layer, where these $m$ filters are nonlearnable and are pointwise monomial transformations of the seed filter $\mathcal{W}_{l}$. The input image $\bm{x}_l$ is filtered by these filters to generate $m$ response maps, which are then passed through a nonlinear activation gate, such as a rectified linear unit (ReLU)~\cite{Nair@Rectified}, and become $m$ feature maps. Accordingly, the process of generating the feature maps can be expressed as,
\begin{align}
    \bm{y} = \sum_{i=1}^m g(f(\bm{w}_i) \ast \bm{x})
\end{align}
\noindent where $g(\cdot)$ is a nonlinear activation, and $f(\bm{w}_i)$ is the monomial filter. 

\begin {figure*}[t]
\centering
\includegraphics[width=0.9\linewidth,clip]{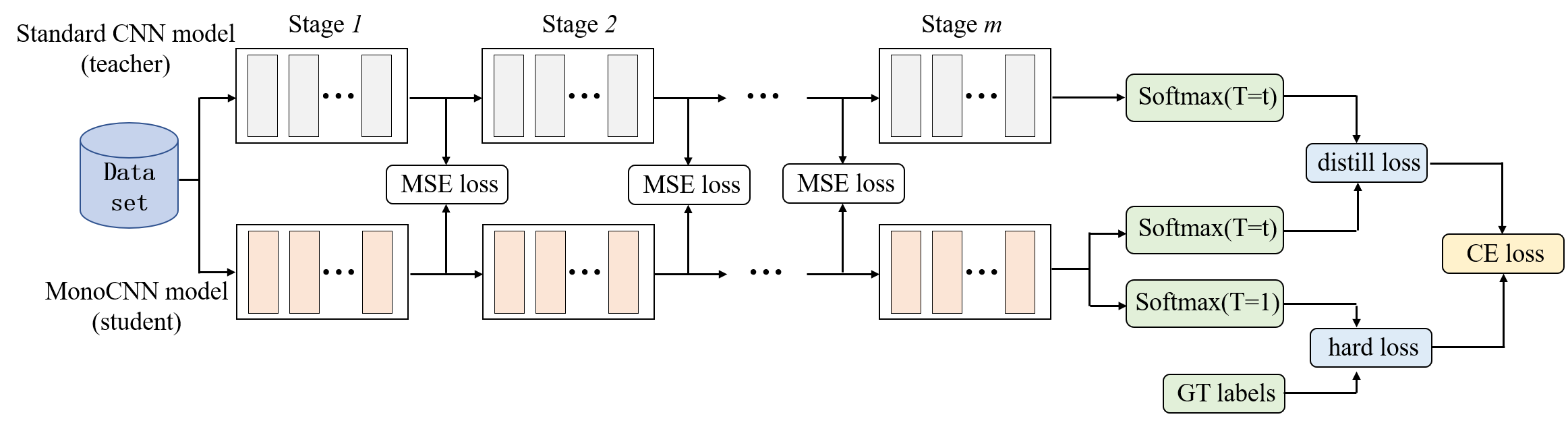} 
\caption {Stagewise supervised training pipeline. Intermediate supervision is imposed between the feature maps of our proposed model and their counterparts.}
\label{fig:stage-KDloss}
\vspace{-0.5cm}
\end{figure*}

Compared to a standard CNN module with the same structure (with $1\times1$ convolutions), the number of learnable parameters is significantly smaller in the \ourmethod{} model. Let us assume that the numbers of input and output channels are $C_{in}$ and $C_{out}$, respectively. Therefore, the size of each 3-D filter in both the CNN and the proposed MonoCNN is $C_{in}\cdot k\cdot k$, where $k$ is the kernel size of the filter, and there are $m$ such filters. The $1\times1$ convolutions act on the $m$ filters and create the $C_{out}$-channel output. For the standard CNN, the number of learnable parameters is $C_{in} \cdot k\cdot k\cdot m + m\cdot C_{out}$. For the \ourmethod{} model, the number of learnable parameters is $C_{in}\cdot k\cdot k \cdot 1 + m\cdot C_{out}$. For simplicity, let us assume that $C_{in}=C_{out}$, which is usually the case for a deep CNN architecture. Then, we have the parameter saving ratio:
{
\begin{align}
\tau = \frac{\textrm{\#P}_{\textrm{CNN}}}{\textrm{\#P}_{\textrm{\ourmethod{}}}} = \frac{C_{in} \cdot k\cdot k\cdot m + m \cdot C_{out}}{C_{in}\cdot k\cdot k \cdot 1 + m\cdot C_{out}} = \frac{k^2 m + m}{k^2 + m} \nonumber
\end{align}
}\noindent 
and when the filter kernel size is $k\!=\!3$ and the number of convolutional filters required for each layer satisfies $m\gg 3^2$, we have a parameter saving ratio of $\tau = \frac{10m}{m+9}\approx10$. It should be mentioned that our proposed \ourmethod{} does not include $1\times1$ convolutions, and thus $m\!=\!C_{in}\!=\!C_{out}$. Consequently, the parameter saving ratio $\tau$ of our proposed \ourmethod{} becomes equal to $m$, i.e., the number of filters per layer in the CNN model; for a high-performance CNN model, there are typically 32, 64, 256, 512, and 1024 filters per layer. Accordingly, our \ourmethod{} achieves parameter savings of approximately $32\times$, $64\times$, $256\times$ or more.

On the cloud server, \ourmethod{} contains only a few learnable parameters while other parameters of the model are randomly generated according to predefined rules and can be saved and reproduced through random seeds. Thus, after the cloud server has trained \ourmethod{}, the cloud server needs to send only the seed filters and the random seeds to the IoT device to reproduce the trained \ourmethod{}. Compared to transmitting all model parameters, sending only seed filters and random seeds can significantly reduce communication costs.

We further explore the use of a stagewise supervised training paradigm to assist in training the \ourmethod{} model. Fig.~\ref{fig:stage-KDloss} depicts the training pipeline. Specifically, given a \ourmethod{} model as a student model, we use its counterpart CNN model (a standard CNN) as a teacher model. We group the network layers into multiple stages, such that feature maps of the same size (i.e., spatial resolution) belong to the same stage while reducing the feature map size by half in each subsequent stage. Let $\bm{z}_i$ denote the output feature maps of the teacher model in the $i$-th stage, and let $\bm{z}_i^{p}$ denote the output feature maps of the student (i.e., \ourmethod{})  model in the $i$-th stage. We use the $\ell_2$-norms between $\bm{z}_i$ and $\bm{z}_i^{p}$ as additional losses to supervise the intermediate feature learning process. In addition, we leverage knowledge distillation (KD)~\cite{Hinton@Distilling}, taking the output probabilities from the teacher model as soft labels. Therefore, the final loss that we backpropagate for training the \ourmethod{} model is defined as follows: 
\begin{align*}
\mathcal{L}(x;\bm{W}) &= \frac{1}{N}\sum_{i=1}^{N}\big\lVert\bm{z}_i - \bm{z}_i^{p}\big\rVert^2_2 \hspace{6em} \mbox{(MSE loss)}\\
& + \ell_{\textrm{CE}}\big(y, q(x;\bm{W})\big) \hspace{7em} \mbox{(hard loss)}\\
& + \ell_{\textrm{CE}}\big(p(x), q(x;\bm{W})\big), \hspace{5em} \mbox{(distill loss)}
\end{align*}
where $x$ and $y$ denote the inputs and outputs, respectively, provided by the dataset; $\bm{W}$ denotes the learnable parameters of the \ourmethod{} model (\ie the seed filter parameters); $p(\cdot)$ and $q(\cdot)$ are the output probabilities of the teacher model and the student (\ie \ourmethod{}) model, respectively; and $\ell_{\textrm{CE}}$ is the cross-entropy loss. Note that we also add $\ell_2$-norms of the learnable parameters to prevent overfitting, which are removed in the above loss formulation for brevity. See Fig.~\ref{fig:stage-KDloss} for a pictorial illustration.  

\begin{figure}[t]
    \centering
    \begin{subfigure}[b]{0.4\textwidth}
        \centering
        \includegraphics[width=.95\textwidth]{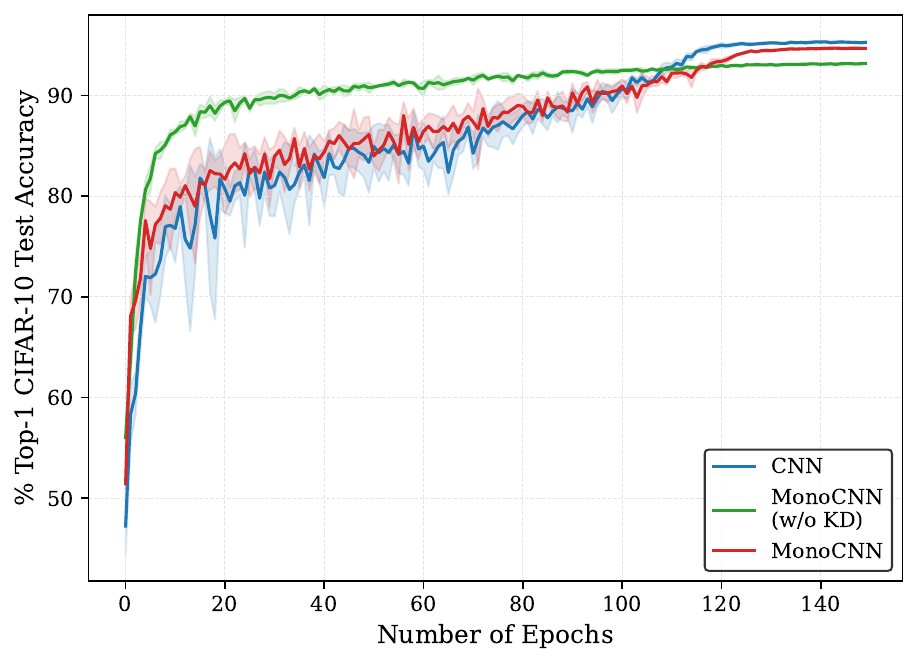}
        \caption{CIFAR-10 data\label{fig:abl_conv_c10}}
    \end{subfigure}\\
    \centering
    \begin{subfigure}[b]{0.4\textwidth}
        \centering
        \includegraphics[width=.95\textwidth]{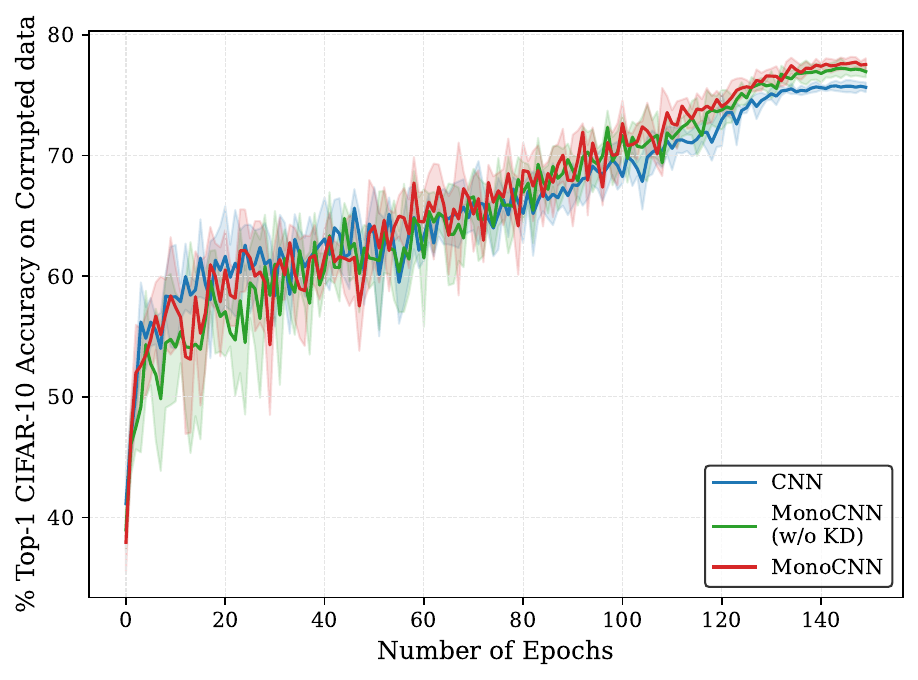}
        \caption{CIFAR-10 corrupted data\label{fig:abl_conv_c10_corrupted}}
    \end{subfigure}
    \caption{Performance and convergence rate of standard CNN, \ourmethod{} (without KD), and \ourmethod{} on the CIFAR dataset (clean vs. corrupted).}\label{fig:performancevsconvergencerate}
    \vspace{-0.5cm}
\end{figure}

Fig.~\ref{fig:performancevsconvergencerate} illustrates the performance and convergence rates of the standard CNN model, the \ourmethod{} model without KD, and the \ourmethod{} model on the CIFAR-10 dataset. As shown in Fig.~\ref{fig:abl_conv_c10}, when processing clean data, the \ourmethod{} model converges faster than the standard CNN model, but the performance is lower. However, when KD is used performance of \ourmethod{} model improves, and its convergence rate decreases, possibly due to the use of a standard CNN as the teacher model. As shown in Fig.~\ref{fig:abl_conv_c10_corrupted}, when processing corrupted data, the three methods achieve comparable convergence rates. The performance of the \ourmethod{} model is higher than that of the standard CNN model and slightly higher than that of the \ourmethod{} model without KD. The experimental results shown in Fig.~\ref{fig:performancevsconvergencerate} demonstrate the superiority of the \ourmethod{} model in handling corrupted data.

\subsection{Filter generation function design} \label{ref:filter generation function}
\begin{figure}[t]
    \centering
    \begin{subfigure}[b]{0.4\textwidth}
        \centering
        \includegraphics[width=.95\textwidth]{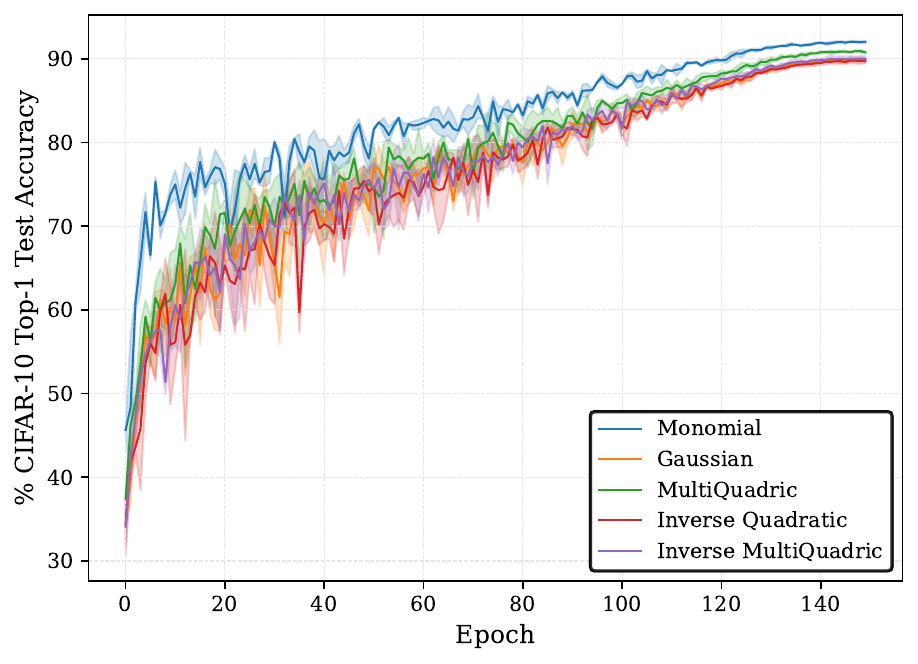}
        \caption{CIFAR-10\label{fig:abl_fgf_c10}}
    \end{subfigure}\hfill
    \centering
    \begin{subfigure}[b]{0.4\textwidth}
        \centering
        \includegraphics[width=.95\textwidth]{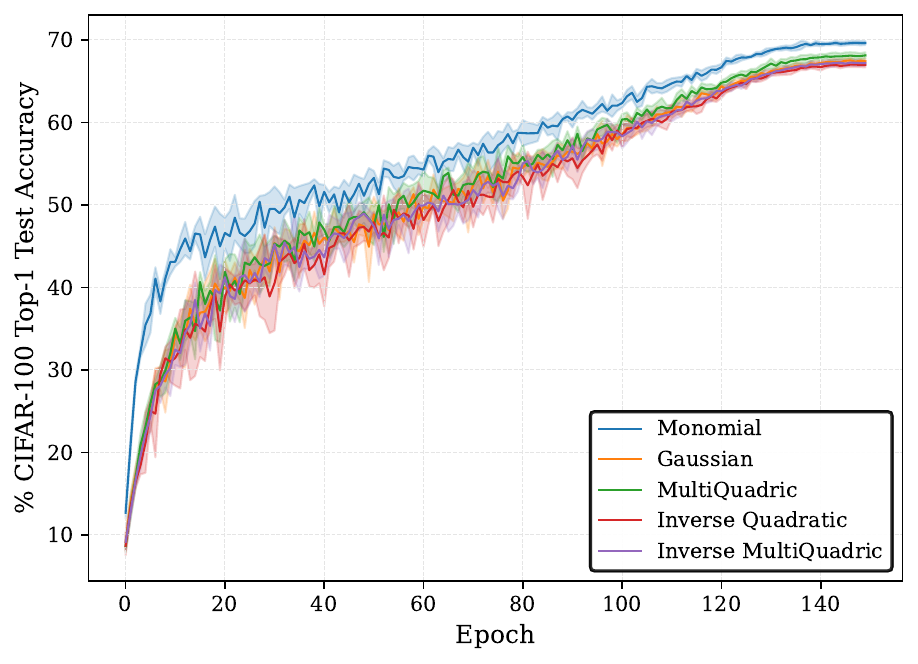}
        \caption{CIFAR-100\label{fig:abl_fgf_c100}}
    \end{subfigure}
    \caption{Performance and convergence rate of five filter generation functions on CIFAR-10 and CIFAR-100 datasets.}
    \label{fig:fivegenerationfunctions}
    \vspace{-0.5cm}
\end{figure}

The combination of using seed filters and generating new filters according to certain rules makes \ourmethod{} comparable to or exceeds the performance of the standard CNN model in handling corrupted data. The rules for generating new filters are of great merit. The standard CNN model contains a large number of nonlinear mappings, which inspired us to use nonlinear mapping functions as FGFs. Given the existing nonlinear mapping functions and the large number of derivations included in the standard CNN model, we choose the following five functions that are easy to compute:
\begin{equation}
\begin{array}{l}
\begin{split}
\varphi (x) = \text{sign} (x)|x|^{\beta}
\end{split},
\end{array}
\label{eq:poly}
\end{equation}

\begin{equation}
\begin{array}{l}
\begin{split}
\varphi (x) = e^{-(\beta x)^2}
\end{split},
\end{array}
\label{eq:gaussian}
\end{equation}

\begin{equation}
\begin{array}{l}
\begin{split}
\varphi (x) = \sqrt{1+(\beta x)^2}
\end{split},
\end{array}
\label{eq:multiquadric}
\end{equation}

\begin{equation}
\begin{array}{l}
\begin{split}
\varphi (x) = \frac{1}{1+(\beta x)^2}
\end{split},
\end{array}
\label{eq:inversequadratic}
\end{equation}

\begin{equation}
\begin{array}{l}
\begin{split}
\varphi (x) =  \frac{1}{\sqrt{1+(\beta x)^2}}
\end{split},
\end{array}
\label{eq:inversemultiquadric}
\end{equation}

Eq.~\ref{eq:poly}, Eq.~\ref{eq:gaussian}, Eq.~\ref{eq:multiquadric}, Eq.~\ref{eq:inversequadratic} and Eq.~\ref{eq:inversemultiquadric} are monomial function, Gaussian function, multiquadric function, inverse quadratic function and inverse multiquadric function, respectively. These functions are nonlinear and easy to compute after derivation. Since a theoretical basis to prove which nonlinear mapping function is the best for generating novel filters proved elusive we empirically evaluate them across different datasets. As shown in Fig.~\ref{fig:fivegenerationfunctions}, and the monomial function is significantly better than the others in terms of performance. Therefore, we use the monomial function in our FGF in this paper and call the CNN model based on the seed filter and monomial function as \ourmethod{}.


\subsection{Discussion}
\subsubsection{Using \ourmethod{} on IoT devices}
After the IoT device receives the seed filters and seeds sent by the cloud server, there are two methods for using MonoCNN.
The first method is to generate the MonoCNN according to the seed filters, seeds and the FGF when the IoT is idle and store it. 
When the MonoCNN model needs to be used, the IoT device can page it into memory to run it in the same way as the standard CNN model.
The second method is to dynamically generate the MonoCNN model.
That is, when the MonoCNN needs to be used, the IoT device instantly generates the MonoCNN by paging the seed filters, seeds, and the FGF into memory.
The second method, which only stores seed filters and seeds on the IoT device, can save memory usage and page-in overhead.
However, the price is that there is a certain overhead in generating the MonoCNN model.
Practically, since the generation process of MonoCNN has only one multiplication and addition operation, its generation overhead is small.
We will test the resource overhead of generating MonoCNN on the IoT device as our future work.


\subsubsection{Theoretical analysis \label{sec:general}}

Here, we provide theoretical analysis on the \ourlayer{} layer and demonstrate how it can well approximate the standard convolutional layer.

At layer $l$, let $\bm{x}_\pi \in \mathds{R}^{(C \cdot k \cdot k) \times 1}$ be a vectorized single patch from the $C$-channel input maps at location $\pi$, where $k$ is the kernel size of the convolutional filter. Let $\bm{w} \in \mathds{R}^{(C \cdot k \cdot k) \times 1}$ be a vectorized single convolution filter from the convolutional filter tensor $\bm{W} \in \mathds{R}^{C\times k \times k \times m}$, which contains a total of $m$ generated convolutional filters at layer $l$. We drop the layer subscription $l$ for brevity.

In a standard CNN, this patch $\bm{x}_\pi$ is taken as a dot product with the filter $\bm{w}$, followed by the nonlinearity (e.g., ReLU $\sigma_{\mathrm{relu}}$), resulting in a single output feature value $d_\pi$ at the corresponding location $\pi$ on the feature map. Similarly, each value of the output feature map is a direct result of convolving the entire input map $\bm{x}$ with a convolutional filter $\bm{w}$. This microscopic process can be expressed as: 
\begin{align}
d_\pi = \sigma_{\mathrm{relu}} ( \bm{w}^{\top} \bm{x}_\pi )
\label{eq:d}
\end{align}
\noindent Without loss of generality, we assume a single-seed \ourlayer{} case for the following analysis. For a \ourlayer{} layer, a single-seed filter $\bm{w}_{s}$ is expanded into a set of $m$ convolutional filters $\bm{W} \in \mathds{R}^{m \times k \times k \times w}$ where $\bm{w}_i = \bm{w}_{s}^{\circ\beta_i}$, and the exponents $\beta_i$s are predefined and are not updated during training. 

The corresponding output feature map value $d_\pi^{\mathrm{~(mono)}}$ from a \ourlayer{} layer is a linear combination of multiple elements from the intermediate response maps. Each slice of this response map is obtained by convolving the input map $\bm{x}$ with $\bm{W}$, followed by a nonlinearity. The corresponding output feature map value $d_\pi^{\mathrm{~(mono)}}$ is thus obtained by linearly combining the $m$ response maps with parameters $\alpha_1, \alpha_2, \ldots, \alpha_m$. This entire process can be expressed as:
\begin{align}
d_\pi^{\mathrm{~(mono)}} = \sigma_{\mathrm{relu}} ( \underbrace{\bm{W}\bm{x}_\pi )^\top}_{1\times m} \underbrace{\boldsymbol{\alpha}}_{m\times1} = \bm{c}_{\mathrm{relu}}^\top \boldsymbol{\alpha}
\label{eq:d_early}
\end{align}
\noindent where $\bm{W}$ is now a 2D matrix of size $m \times k^2w$ with $m$ filters $\mathrm{vec}({\bm{w}_i})$ stacked as rows, with a slight abuse of notation. 
$\boldsymbol{\alpha} = [\alpha_1,\ldots,\alpha_m]^\top \in \mathds{R}^{m\times1}$.
Comparing $d_\pi$ and $d_\pi^{\mathrm{~(mono)}}$, we consider the following two cases (i) $d_\pi=0$: since $\bm{c}_{\mathrm{relu}} = \sigma_{\mathrm{relu}} ( \bm{W} \bm{x}_\pi ) \geq 0$, there always exists a vector $\boldsymbol{\alpha} \in \mathds{R}^{m\times1}$ such that $d_\pi^{\mathrm{~(mono)}}=d_\pi$. 
However, when (ii) $d_\pi>0$, it is obvious that the approximation does not hold when $\bm{c}_{\mathrm{relu}}=\bm{0}$. 
Thus, under the assumption that $\bm{c}_{\mathrm{relu}}$ is not an all-zero vector, the approximation $d_\pi^{\mathrm{~(mono)}} \approx d_\pi$ will hold.

\begin{figure*}[ht]
    \centering
    \includegraphics[width=.96\textwidth]{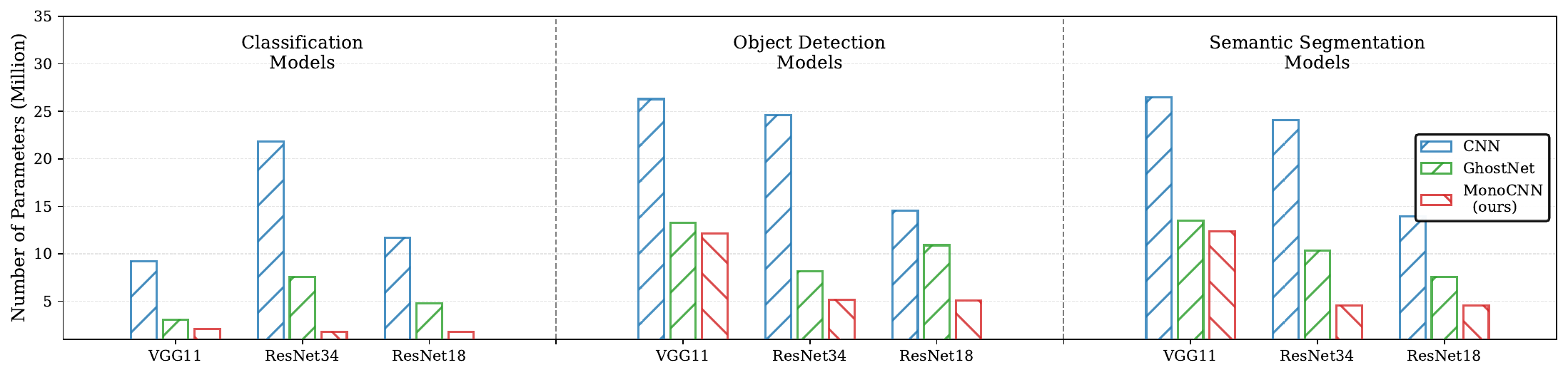}
    \caption{Comparison of the number of parameters required to be transmitted by the regular CNN and our \ourmethod{}. We consider three widely used architectures and three important vision tasks. We also include GhostNet as a reference for competitor methods.\label{fig:network_traffic}}
    \vspace{-0.2cm}
\end{figure*}

\begin{table*}[ht]
\centering
\caption{Comparing \ourmethod{} and existing alternative methods for image classification on CIFAR-10/-100. Mean performance along with standard deviation from five runs are reported (i.e., mean$_{\pm\mbox{\scriptsize std}}$). \label{tab:cifar}}
\centering
\begin{subtable}[b]{0.32\textwidth}
\centering
\resizebox{.9\textwidth}{!}{%
    \begin{tabular}{@{\hspace{2mm}}lccc@{\hspace{2mm}}}
    \toprule
    \multirow{2}{*}{Method} & \multirow{2}{*}{\begin{tabular}[c]{@{}c@{}}Params \\ (M)\end{tabular}} & \multicolumn{2}{c}{Top-1 Accuracy (\%)} \\ \cmidrule(l){3-4} 
     &  & CIFAR-10 & CIFAR-100 \\ \midrule
    CNN & 9.2 & 92.44\textcolor{darkgray}{$_{\pm0.19}$} & 71.69\textcolor{darkgray}{$_{\pm0.10}$} \\
    LBCNN & 1.2 & 87.67\textcolor{darkgray}{$_{\pm0.30}$} & 60.57\textcolor{darkgray}{$_{\pm0.10}$} \\
    PNN & 1.2 & 70.61\textcolor{darkgray}{$_{\pm0.17}$} & 43.56\textcolor{darkgray}{$_{\pm0.10}$} \\
    ShiftNet & 1.1 & 30.83\textcolor{darkgray}{$_{\pm0.88}$} & 8.84\textcolor{darkgray}{$_{\pm0.46}$} \\
    GhostNet & 4.9 & 90.05\textcolor{darkgray}{$_{\pm0.25}$} & 65.59\textcolor{darkgray}{$_{\pm0.19}$} \\ \midrule
    \ourmethod{} & 2.4 & 91.45\textcolor{darkgray}{$_{\pm0.15}$} & 69.17\textcolor{darkgray}{$_{\pm0.21}$} \\ \bottomrule
    \end{tabular}%
}
\caption{{\footnotesize VGG 11}\label{tab:cifar_vgg}}
\end{subtable} \hfill
\centering
\begin{subtable}[b]{0.32\textwidth}
\centering
\resizebox{.9\textwidth}{!}{%
    \begin{tabular}{@{\hspace{2mm}}lccc@{\hspace{2mm}}}
    \toprule
    \multirow{2}{*}{Method} & \multirow{2}{*}{\begin{tabular}[c]{@{}c@{}}Params \\ (M)\end{tabular}} & \multicolumn{2}{c}{Top-1 Accuracy (\%)} \\ \cmidrule(l){3-4} 
     &  & CIFAR-10 & CIFAR-100 \\ \midrule
    CNN & 11.2 & 95.19\textcolor{darkgray}{$_{\pm0.11}$} & 77.98\textcolor{darkgray}{$_{\pm0.33}$} \\
    LBCNN & 2.8 & 93.05\textcolor{darkgray}{$_{\pm0.10}$} & 72.72\textcolor{darkgray}{$_{\pm0.02}$} \\
    PNN & 2.8 & 92.45\textcolor{darkgray}{$_{\pm0.18}$} & 73.21\textcolor{darkgray}{$_{\pm0.13}$} \\
    ShiftNet & 2.8 & 92.76\textcolor{darkgray}{$_{\pm0.14}$} & 73.67\textcolor{darkgray}{$_{\pm0.24}$} \\
    GhostNet & 4.3 & 93.40\textcolor{darkgray}{$_{\pm0.14}$} & 72.77\textcolor{darkgray}{$_{\pm0.58}$} \\ \midrule
    \ourmethod{} & 2.8 & 94.02\textcolor{darkgray}{$_{\pm0.06}$} & 74.22\textcolor{darkgray}{$_{\pm0.13}$} \\ \bottomrule
    \end{tabular}%
}
\caption{\footnotesize ResNet18 \label{tab:cifar_resnet}}
\end{subtable} \hfill
\centering
\begin{subtable}[b]{0.32\textwidth}
\centering
\resizebox{.9\textwidth}{!}{%
    \begin{tabular}{@{\hspace{2mm}}lccc@{\hspace{2mm}}}
    \toprule
    \multirow{2}{*}{Method} & \multirow{2}{*}{\begin{tabular}[c]{@{}c@{}}Params \\ (M)\end{tabular}} & \multicolumn{2}{c}{Top-1 Accuracy (\%)} \\ \cmidrule(l){3-4} 
     &  & CIFAR-10 & CIFAR-100 \\ \midrule
    CNN & 21.3 & 95.57\textcolor{darkgray}{$_{\pm0.08}$} & 78.73\textcolor{darkgray}{$_{\pm0.42}$} \\
    LBCNN & 3.9 & 93.54\textcolor{darkgray}{$_{\pm0.16}$} & 73.81\textcolor{darkgray}{$_{\pm0.21}$} \\
    PNN & 3.9 & 92.31\textcolor{darkgray}{$_{\pm0.20}$} & 73.33\textcolor{darkgray}{$_{\pm0.14}$} \\
    ShiftNet & 3.9 & 92.84\textcolor{darkgray}{$_{\pm0.20}$} & 73.87\textcolor{darkgray}{$_{\pm0.25}$} \\
    GhostNet & 7.1 & 93.58\textcolor{darkgray}{$_{\pm0.24}$} & 73.16\textcolor{darkgray}{$_{\pm0.63}$} \\ \midrule
    \ourmethod{} & 4.0 & 94.24\textcolor{darkgray}{$_{\pm0.12}$} & 75.63\textcolor{darkgray}{$_{\pm0.52}$} \\ \bottomrule
    \end{tabular}%
}
\caption{\footnotesize ResNet34 \label{tab:cifar_mobilenet}}
\end{subtable}
\end{table*}

\section{Evaluation} \label{ref:4-experiment} 
In this section, we first introduce our experimental setup including the datasets, baselines, and evaluation metrics studied in this work, followed by the implementation details. We then provide an empirical comparison in terms of network complexity and performance on multiple vision benchmarks.    
\subsection{Experimental Setup}
\noindent\textbf{Datasets.} Five popular datasets are used to verify the effectiveness of the proposed method.

{\bf{CIFAR-10/-100}}~\cite{cifar10} are two multiclass natural object datasets widely used for image classification. Both consist of 50,000 training and 10,000 test images from 10/100 classes, with each image of $32\times32$ pixels.
    
{\bf{MS COCO}}~ \cite{mscoco} dataset comprises more than 100K images of diverse objects with annotations, including both bounding boxes and segmentation masks, from 80 categories. We take the \emph{train2017} set for training and compare detection performance on the \emph{val2017} set. 
    
{\bf{PASCAL VOC 2012}}~\cite{pascal-voc} is a comparably small-scale dataset of images with 20 foreground object categories and one category for background. Following prior works \cite{liu2019auto}, we augment the original training set with the extra annotations from \cite{6126343}, resulting in 10, 582 images (\emph{train\_aug}) in total for training. We use this dataset for both object detection and semantic segmentation. 
    
{\bf{Cityscapes}}~\cite{cityscapes} is a large-scale (images are of $1024\times2048$ pixels) dataset for semantic understanding of urban street scenes. It is officially split into a training set of 2, 975 images, a validation set of 500 images, and a (privately hosted) testing set of 1, 525 images. We use 19 from the provided 30 classes for semantic segmentation.

\noindent\textbf{Baselines.} To verify the effectiveness of the proposed method, we consider the following baselines:

{\bf{LBCNN}}~\cite{juefei2017local}: The local binary convolutional neural network (LBCNN) uses sparse local binary filter parameters (randomly initialized and kept fixed) followed by a learned $1\times1$ convolution to replace regular $3\times3$ convolution layers. 

{\bf{PNN}}~\cite{juefei2018perturbative}: The perturbative neural network (PNN) injects randomly generated additive noise to the input features combined through a learned $1\times1$ convolution to replace regular $3\times3$ convolution layers. 

{\bf{ShiftNet}}~\cite{chen2019all}: ShiftNet applies a sparse spatial shift (e.g., one pixel left) to create diverse viewpoints of features, replacing the regular $3\times3$ convolutions. 

{\bf{GhostNet}}~\cite{han2020ghostnet}: GhostNet partially substitutes computationally expensive operations (e.g., regular $3\times3$ convolutions) with cheap operations (e.g., $1\times1$ or grouped $3\times3$ convolutions). 

To ensure a fair and comprehensive comparison, we implement all the above baseline methods within three well-studied underlining architectures, including VGG11 \cite{Simonyan@Very}, ResNet18 \cite{He@Deep}, and ResNet34 \cite{He@Deep}. 

\vspace{3pt}
\noindent\textbf{Evaluation Metrics.}
We use top-1 accuracy to compare performance for image classification. We use the mean average precision (AP), computed for a recall value over 0 to 1, for object detection. For semantic segmentation, we adopt mean intersection-over-union (mIoU), which computes the IoU for each semantic class averaged over classes. It is worth noting that we only consider the number of parameters that must be learned, as these are the parameters that must be sent from the cloud server to IoT devices. 

\vspace{3pt}
\noindent\textbf{Implementation Details.}
We implement our method in PyTorch 1.7 with CUDA 10.1, and all experiments are performed on 2080TI GPUs. Following the suggestions from the original papers, we set the sparsity to 0.9 for LBCNN~\cite{juefei2017local} and the noise level to 0.01 for PNN~\cite{juefei2018perturbative}; we use the $1\times1$ convolution as the cheap operation for GhostNet~\cite{han2020ghostnet} and set the ratio to 4. 

\subsection{Experimental Results}
In this section, we first present a comparison of network complexity, followed by a performance comparison for image classification, object detection, and semantic segmentation on clean data. Finally, we compare robustness on limited training data, corrupted data, and different style data. 

\subsubsection{Amount of Model Parameter Transmission}
Our proposed \ourmethod{} minimizes the number of model parameters sent by the cloud server to IoT devices.
As shown in Fig.~\ref{fig:network_traffic}, we consider three widely used architectures (VGG11 and ResNet18/34) and compare the learnable parameters of \ourmethod{} with those of regular CNNs for image classification, object detection, and semantic segmentation.
Since all filter parameters in the standard CNN model need to be learned, the cloud server needs to send all the filter parameters of the standard CNN model to the IoT device, resulting in a large amount of model parameter transmission.
In contrast, in our proposed \ourmethod{}, only a single-seed filter needs to be learned in each layer, and the rest of the filters are generated by the filter generation function.
The hyperparameters of the filter generation function (e.g., monomial exponent) are randomly initialized and remain fixed so that these nonlearnable hyperparameters can be saved and reproduced by the random number generator seed.
Therefore, cloud-assisted training of \ourmethod{} only requires the cloud server to send a few seed filters and the random number generator seeds to recover the \ourmethod{} model on the IoT device.

Additionally, as shown in Fig.~\ref{fig:network_traffic}, GhostNet also has fewer model parameters than the standard CNN model because GhostNet uses cheap operations to augment filters. 
However, our proposed \ourmethod{} needs to send fewer model parameters, and in subsequent experiments, our proposed \ourmethod{} outperforms GhostNet in almost all tasks. 
It is worth mentioning that other types of parameter reduction techniques (e.g., pruning, quantization \cite{Han@Deep}, and neural architecture search \cite{nat,9201169}) can be applied on top of our method for further compression of model parameters.

\subsubsection{Results on Standard Benchmarks}
In this section, we evaluate the effectiveness of our \ourmethod{} on standard benchmark datasets for image classification, object detection, and semantic segmentation tasks. 

\vspace{3pt}
\noindent\textbf{Image Classification.}
For training on the CIFAR-10/-100 datasets, we use the SGD optimizer with an initial learning rate of 0.025, which is annealed to zero following the cosine schedule. We use standard data augmentations: we pad images with four pixels on each side and randomly crop a $32\times32$ region, from which random horizontal flipping is also applied. Given the stochastic nature of the CIFAR datasets (as the results are subject to high variance even with exactly the same setup), we repeat the training five times with different initial random seeds and report the mean performance along with the standard deviation.

Table~\ref{tab:cifar} depicts the results. In general, we observe that our \ourmethod{} consistently outperforms other peer methods on both CIFAR-10 and CIFAR-100 while requiring a similar or fewer number of parameters to be learned. Additionally, our \ourmethod{} provides substantial savings in the parameters while achieving similar accuracy performance when compared to regular CNNs. In particular, the proposed \ourmethod{} is \textbf{3.58\% more accurate} on CIFAR-100 and \textbf{2$\bm{\times}$ more compact} than GhostNet \cite{han2020ghostnet} when paired with the VGG11 architecture.

\vspace{3pt}
\noindent\textbf{Objection Detection.}
To evaluate the effectiveness of our model for object detection, we implement all compared methods using ResNet18 as the underlining backbone architecture and FPN \cite{lin2017feature} as the detection head. For training on both MS COCO and PASCAL VOC 2012, we use the SGD optimizer with an initial learning rate of 0.02 and a batch size of eight over four GPU cards. Following the common practice, we adopt the 1$\times$ (i.e., 12 or 36 epochs) schedule to train our detection models and decay the learning rate at the 8th and 11th epochs by a factor of 10. We resize the training images to the shorter side of 800 pixels with the longer side to be within 1333 pixels for MS COCO. We resize the training images to $1000\times600$ for PASCAL VOC 2012.  

\begin{figure*}[ht]
    \centering
    \includegraphics[width=1\textwidth]{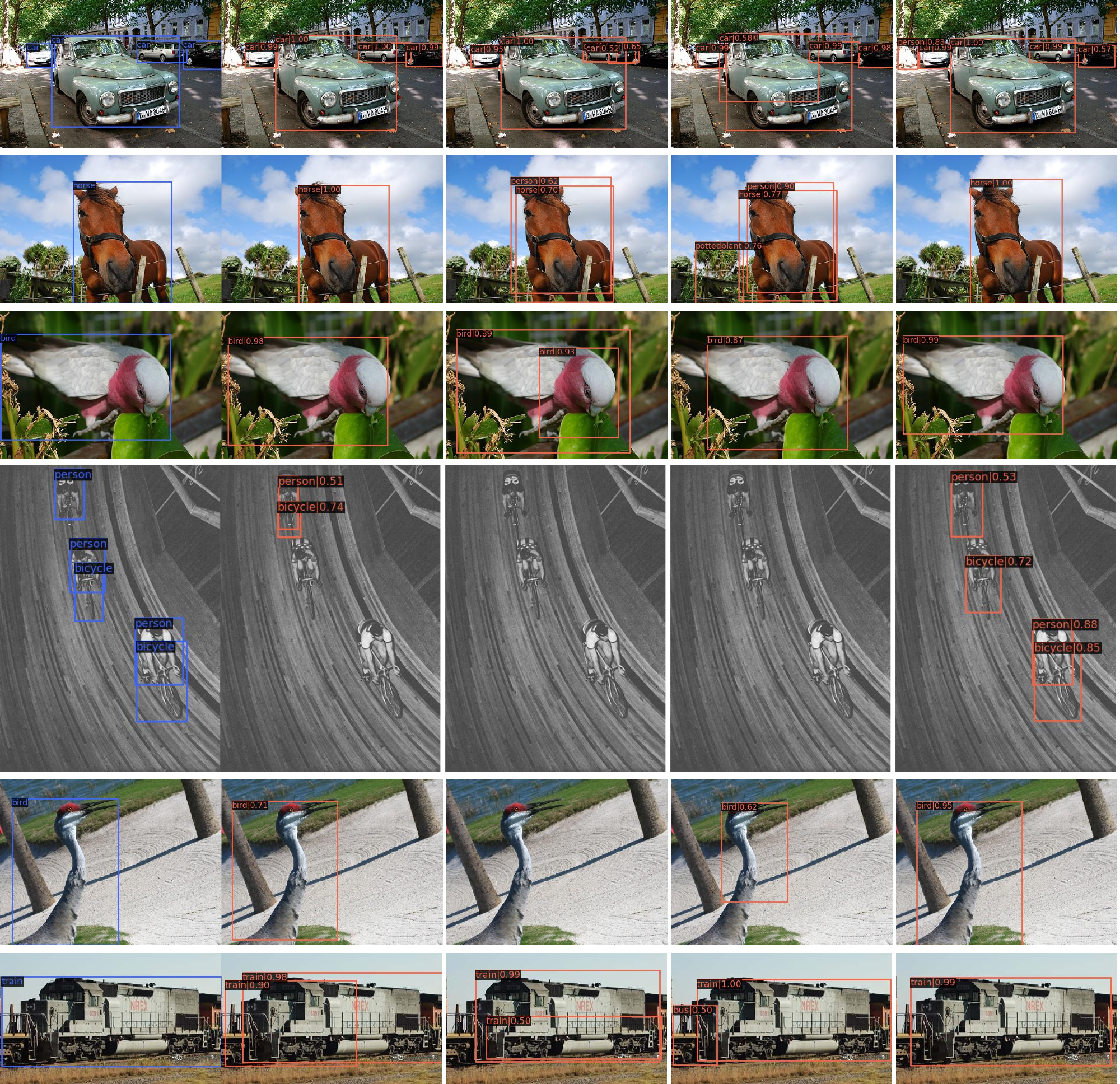}
    \caption{Qualitative comparison on MS COCO object detection. From left to right, we show the example predictions from ground truth, regular CNN, LBCNN, GhostNet, and our \ourmethod{}. The predicted labels with confidence scores are annotated at the top-left corners of the detection boxes. \label{fig:mscoco_vis}}
\end{figure*}

Table~\ref{tab:det_mscoco} and Table~\ref{tab:det_pascal} depict the results. Similar to the previous case of image classification, the proposed \ourmethod{} consistently outperforms other peer methods for object detection. In particular, \ourmethod{} achieves \textbf{6.2 and 8.0 higher AP points} than LBCNN \cite{juefei2017local} while using a similar number of parameters. In addition, we also provide a qualitative visualization between \ourmethod{} and the compared methods in Fig.~\ref{fig:mscoco_vis}. Evidently, \ourmethod{} (right-most column in Fig.~\ref{fig:mscoco_vis}) is not only more accurate in detecting smaller objects (see the first and fourth row in Fig.~\ref{fig:mscoco_vis}) but also more precise in avoiding duplicate detection boxes (see second and third row in Fig.~\ref{fig:mscoco_vis}) than peer methods (Columns 2-4 in Fig.~\ref{fig:mscoco_vis}). 

\begin{table}[h]
\centering
\caption{Comparing \ourmethod{} and existing alternative methods for object detection on MS COCO. \label{tab:det_mscoco}}
\resizebox{.4\textwidth}{!}{%
\begin{tabular}{@{\hspace{2mm}}l|c|cc|ccc@{\hspace{2mm}}}
\toprule
Method & AP & AP$_{50}$ & AP$_{75}$ & AP$_{s}$ & AP$_{m}$ & AP$_{l}$ \\ \midrule
CNN & 33.1 & 52.6 & 35.5 & 18.9 & 35.4 & 43.1 \\
LBCNN & 25.6 & 43.3 & 26.3 & 13.4 & 27.2 & 34.2 \\
GhostNet & 30.4 & 49.3 & 32.1 & 16.8 & 32.5 & 40.8 \\ \midrule
\ourmethod{} & 31.8 & 51.3 & 34.1 & 17.3 & 33.9 & 41.9 \\ \bottomrule
\end{tabular}%
}
\end{table}

\begin{table}[ht]
\centering
\caption{Comparing \ourmethod{} and existing alternative methods for object detection on PASCAL VOC 2012. \label{tab:det_pascal}}
\resizebox{.4\textwidth}{!}{%
\begin{tabular}{@{\hspace{2mm}}l|c|cc|ccc@{\hspace{2mm}}}
\toprule
Method & AP & AP$_{50}$ & AP$_{75}$ & AP$_{s}$ & AP$_{m}$ & AP$_{l}$ \\ \midrule
CNN & 47.4 & 79.8 & 50.6 & 19.1 & 33.9 & 52.1 \\
LBCNN & 37.7 & 69.5 & 36.1 & 18.0 & 25.1 & 41.8 \\
GhostNet & 44.3 & 76.7 & 45.9 & 17.2 & 31.1 & 48.8 \\ \midrule
\ourmethod{} & 45.7 & 78.1 & 47.7 & 16.5 & 30.8 & 50.8 \\ \bottomrule
\end{tabular}%
}
\end{table}

\vspace{3pt}
\noindent\textbf{Semantic Segmentation.} We follow the same setup as in the previous case of object detection. We also implement all compared methods using ResNet18 as the underlining backbone architecture and FPN as the segmentation head. For training on Cityscapes and PASCAL VOC 2012, we use the SGD optimizer with a momentum of 0.9 and weight decay of {5e-4}. The batch size is set to 24 over two 2080TI GPUs. Following the common practice, we adopt the ``poly'' learning rate policy (i.e., $0.01 \times (1 - \frac{iter}{maxIter})^{0.9}$) from 0.01 to zero in 60K iterations. Data augmentation includes color jittering, random horizontal flipping, random cropping and random resizing. In addition, we scale training images with a factor randomly sampled from [0.125, 1.5] and crop them to $1024\times512$ for Cityscapes.

\begin{figure*}[ht]
    \centering
    \begin{subfigure}[b]{0.545\textwidth}
    \centering
    \includegraphics[width=\textwidth]{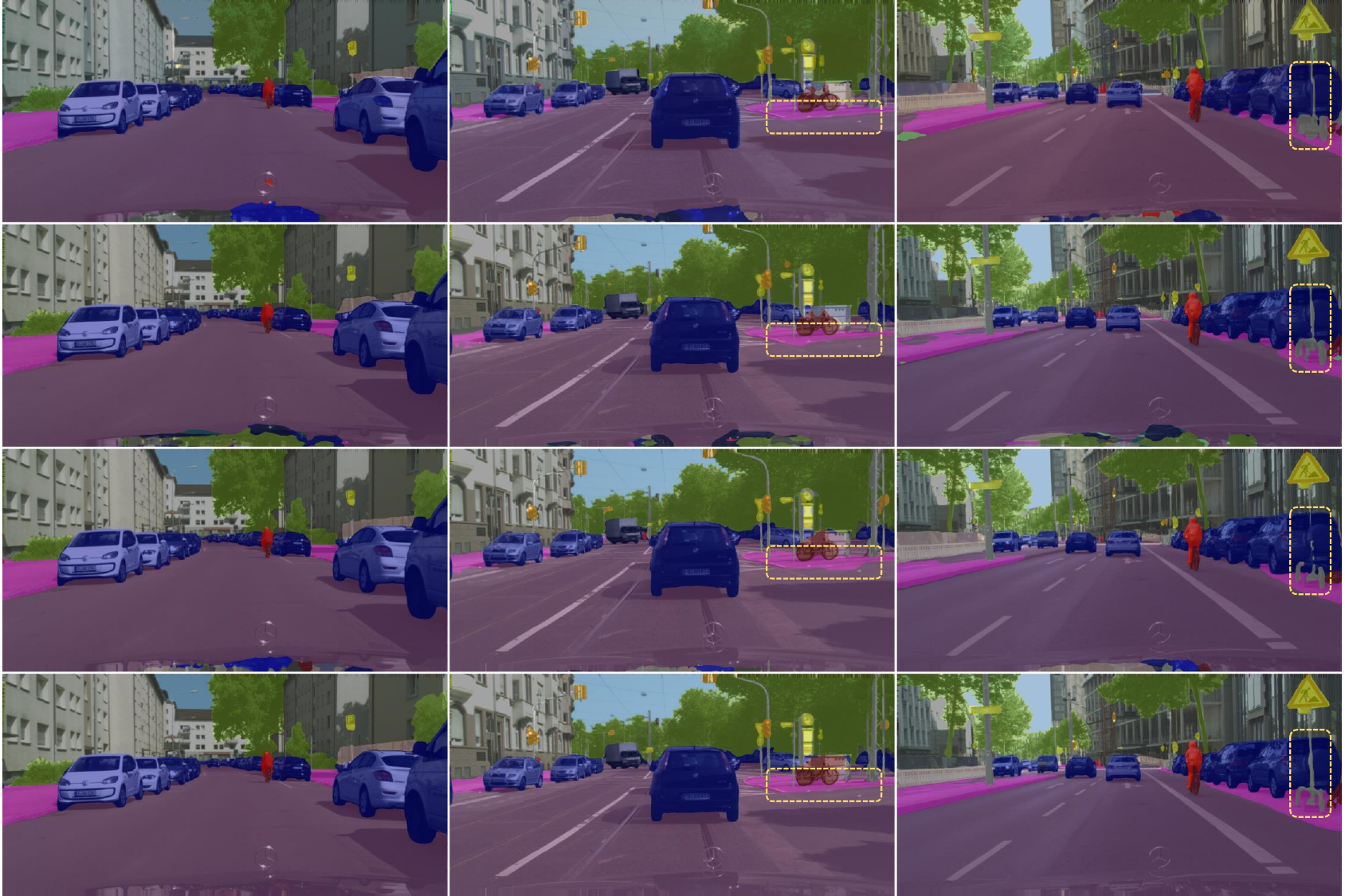}
    \caption{Cityscapes \label{fig:seg_city_vis}}
    \end{subfigure} \hfill
    \centering
    \begin{subfigure}[b]{0.438\textwidth}
    \centering
    \includegraphics[width=\textwidth]{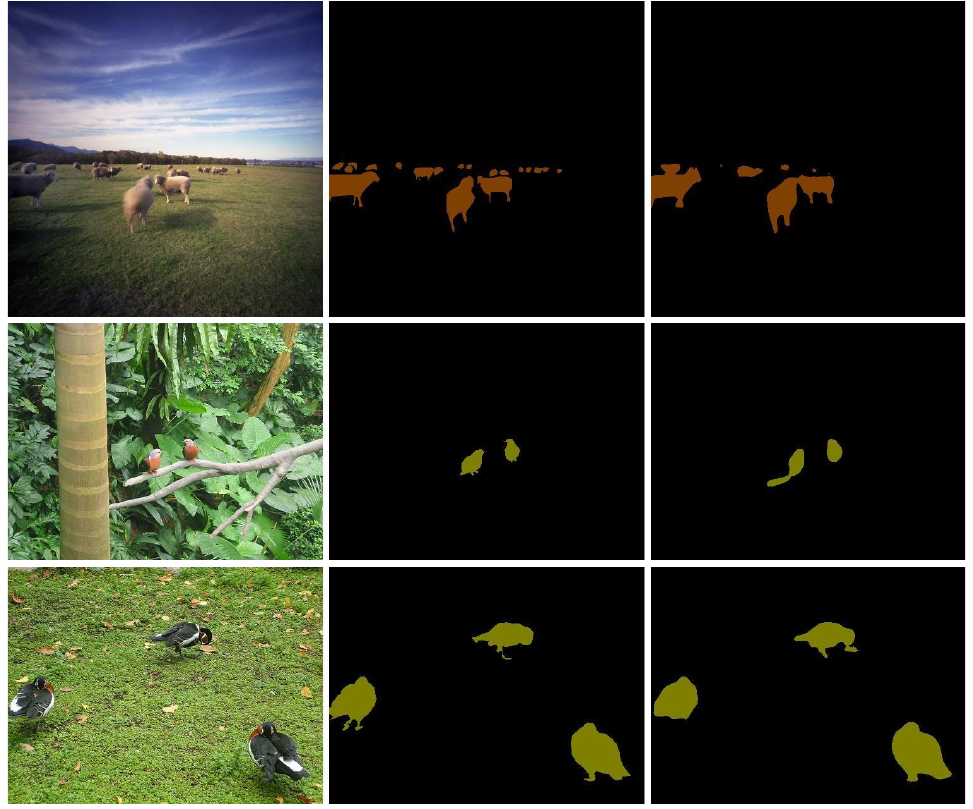}
    \caption{PASCAL VOC 2012}
    \end{subfigure}
    \caption{Qualitative comparison on semantic segmentation. For (a) Cityscapes, we visualize the ground truth, LBCNN, GhostNet, and our \ourmethod{} from top to bottom. For (b) PASCAL VOC 2012, we visualize input images, ground truth, and our \ourmethod{} from left to right. Zoom in for details. \label{fig:seg_vis}}
\end{figure*}

Table~\ref{tab:seg_pascal} and Table~\ref{tab:seg_cityscapes} break down the classwise segmentation mIoU for PASCAL VOC 2012 and Cityscapes, respectively. Evidently, we observe that our \ourmethod{} significantly outperforms peer competitors on both datasets. In particular, \ourmethod{} achieves \textbf{better mIoU with 3$\bm{\times}$ fewer parameters} than the regular CNN model on PASCAL VOC 2012; \ourmethod{} achieves \textbf{3.6 and 5.2 points higher mIoU} than LBCNN \cite{juefei2017local} on the two datasets, respectively. A qualitative comparison is also provided in Fig.~\ref{fig:seg_vis}. Visually, we observe that \ourmethod{} leads to a more fain-grained segmentation on small objects (see boxed regions in \ref{fig:seg_city_vis}). 

\begin{table*}[ht]
\centering
\caption{Comparing \ourmethod{} and existing alternative methods for semantic segmentation on PASCAL VOC 2012. \label{tab:seg_pascal}}
\resizebox{.98\textwidth}{!}{%
\begin{tabular}{@{\hspace{2mm}}l|cccccccccccccccccccc|c@{\hspace{2mm}}}
\toprule
Method & aero & bike & bird & boat & bottle & bus & car & cat & chair & cow & table & dog & horse & mbike & person & plant & sheep & sofa & train & tv & mIoU \\ \midrule
CNN & 79.6 & 37.4 & 71.4 & 51.3 & 54.2 & 81.5 & 76.5 & 77.6 & 28.1 & 59.9 & 38.3 & 67.8 & 66.6 & 71.7 & 78.4 & 39.2 & 66.4 & 34.4 & 69.7 & 60.4 & 65.2 \\
LBCNN & 71.8 & 35.0 & 58.8 & 47.0 & 47.1 & 76.4 & 73.0 & 72.4 & 21.4 & 46.9 & 37.8 & 59.5 & 51.4 & 63.8 & 71.4 & 31.5 & 64.0 & 30.5 & 65.4 & 52.3 & 61.9 \\
GhostNet & 77.4 & 36.2 & 69.3 & 48.8 & 56.4 & 78.4 & 74.7 & 75.8 & 27.1 & 59.1 & 40.8 & 66.0 & 62.6 & 68.5 & 76.0 & 36.3 & 66.5 & 31.3 & 68.4 & 58.8 & 63.5 \\ \midrule
\ourmethod{} & 83.9 & 37.8 & 78.2 & 53.3 & 58.8 & 89.5 & 77.9 & 82.8 & 31.2 & 58.9 & 38.3 & 71.6 & 71.8 & 75.0 & 77.6 & 48.2 & 73.9 & 35.2 & 77.4 & 63.7 & 65.5 \\ \bottomrule
\end{tabular}%
}
\end{table*}

\begin{table*}[ht]
\centering
\caption{Comparing \ourmethod{} and existing alternative methods for semantic segmentation on Cityscapes.\label{tab:seg_cityscapes}}
\resizebox{.98\textwidth}{!}{%
\begin{tabular}{@{\hspace{2mm}}l|ccccccccccccccccccc|c@{\hspace{2mm}}}
\toprule
Method & road & sidewalk & building & wall & fence & pole & light & sign & vegetation & terrain & sky & person & rider & truck & bus & caravan & trailer & train & motorcycle & mIoU \\ \midrule
CNN & 97.3 & 78.8 & 89.9 & 50.1 & 47.4 & 47.4 & 55.5 & 66.8 & 89.9 & 58.6 & 93.0 & 72.7 & 50.3 & 92.4 & 63.3 & 74.0 & 53.4 & 51.0 & 67.9 & 68.4 \\
LBCNN & 96.9 & 76.4 & 88.2 & 47.8 & 42.3 & 40.6 & 44.2 & 59.9 & 88.6 & 56.3 & 92.1 & 66.8 & 42.2 & 90.6 & 56.2 & 65.2 & 36.1 & 41.6 & 63.6 & 62.9\\
GhostNet & 97.3 & 79.4 & 89.2 & 47.9 & 46.5 & 45.7 & 51.7 & 64.7 & 89.6 & 59.9 & 92.6 & 70.7 & 46.6 & 91.5 & 60.0 & 75.8 & 65.7 & 41.4 & 66.2 & 67.5\\ \midrule
\ourmethod{} & 97.4 & 79.8 & 89.7 & 49.0 & 48.7 & 46.0 & 55.0 & 65.7 & 90.0 & 61.0 & 93.0 & 71.3 & 49.1 & 92.1 & 66.4 & 73.0 & 58.9 & 39.4 & 66.5 & 68.1 \\ \bottomrule
\end{tabular}%
}
\end{table*}

\vspace{3pt}
\noindent\textbf{Discussion.}
As shown by experimental results on image classification, object detection, and semantic segmentation tasks, the proposed \ourmethod{} consistently outperforms a wide range of existing alternatives with similar or fewer parameters. 
In addition, the proposed \ourmethod{} can significantly decrease the number of parameters compared to standard CNN models, but with slight performance degradation.
The main reason is that it is difficult for \ourmethod{} with a small number of learnable parameters to process the test images that are highly correlated with the training images through the training images.
However, in real scenarios, in the data collected by IoT devices, the correlation between training images and test images is much smaller than that of the training image set and test image set divided by the standard dataset. 
For \ourmethod{}, its parameters are also affected by the filter generation function and are not completely dependent on the training data, thus \ourmethod{} is expected to achieve high performance in processing this type of image.

\begin{figure}[ht]
    \centering
    \includegraphics[width=.49\textwidth]{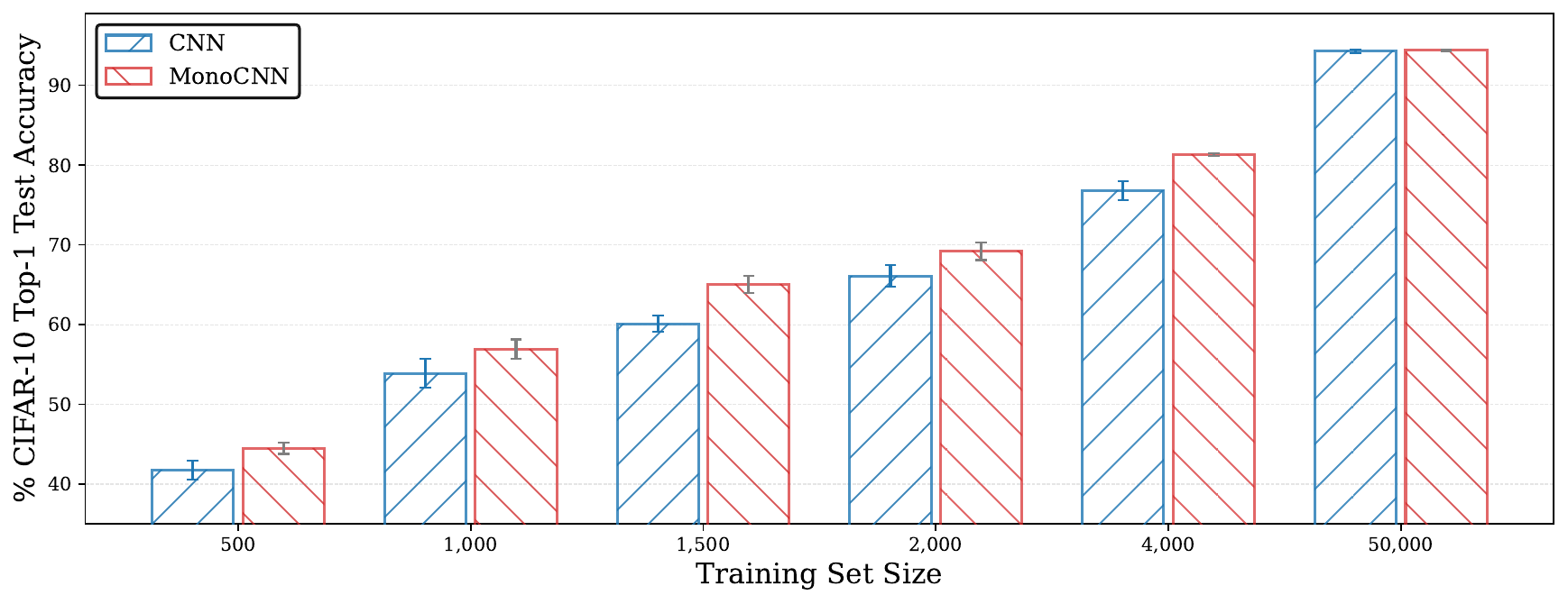}
    \caption{Robustness to limited training data. \label{fig:limited_data_robustness}}
\end{figure}

\subsubsection{Results on Robustness}
In this section, we use CIFAR-10 classification to evaluate the performance of the proposed \ourmethod{} for robustness on limited data and on data with commonly observable corruptions. 

\vspace{3pt}
\noindent\textbf{Limited training data.}
Insufficient training data are a conventional difficulty for deep neural network models but often arise in practical applications. Considering the lower model complexity (i.e., fewer learnable parameters), we hypothesize that \ourmethod{} may be less prone to overfitting to the limited training data. To verify this hypothesis, we perform an empirical experiment on (randomly selected) subsets of the CIFAR-10 training set while keeping the testing set intact. Fig.~\ref{fig:limited_data_robustness} depicts the results. Compared to fully learned convolutions (standard CNNs), \ourmethod{} exhibits noticeably better generalization performance under limited training data.

\begin{table}[ht]
    \centering
    \caption{Details of the corruption types evaluated.\label{tab:corruptions}}
    \resizebox{.45\textwidth}{!}{%
    \begin{tabular}{@{\hspace{2mm}}l|l@{\hspace{2mm}}}
    \toprule
    \multicolumn{1}{l|}{\textbf{Group}} & \multicolumn{1}{c}{\textbf{Corruption Types}} \\ \midrule
    Noise & \begin{tabular}[c]{@{}l@{}}Gaussian, Impulse, Shot, Speckle\end{tabular} \\ \midrule
    Blur & \begin{tabular}[c]{@{}l@{}}Defocus, Glass, Motion, Zoom, Gaussian\end{tabular} \\ \midrule
    Weather & \begin{tabular}[c]{@{}l@{}}Brightness, Fog, Frost, Snow, Spatter\end{tabular} \\ \midrule
    Digital & Contrast, Elastic, JPEG compression, Pixelate, Saturate \\
    \bottomrule
    \end{tabular}%
    }
\end{table}

\begin{figure*}[ht]
    \centering
    \begin{subfigure}[b]{0.192\textwidth}
    \centering
    \includegraphics[width=\textwidth]{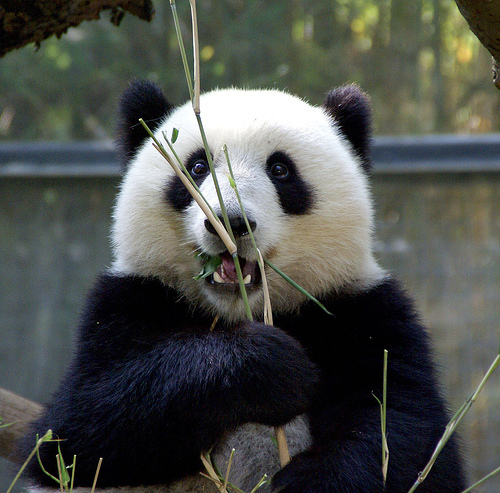}
    \caption{Original}
    \end{subfigure}\hfill
    \centering
    \begin{subfigure}[b]{0.19\textwidth}
    \centering
    \includegraphics[width=\textwidth]{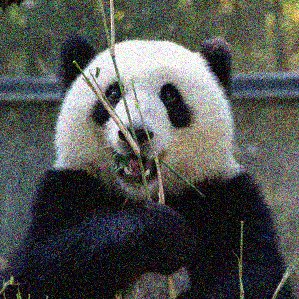}
    \caption{Gaussian noise}
    \end{subfigure}\hfill
    \centering
    \begin{subfigure}[b]{0.19\textwidth}
    \centering
    \includegraphics[width=\textwidth]{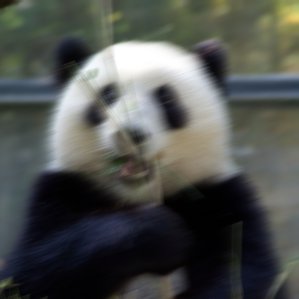}
    \caption{Motion blur}
    \end{subfigure}\hfill
    \centering
    \begin{subfigure}[b]{0.19\textwidth}
    \centering
    \includegraphics[width=\textwidth]{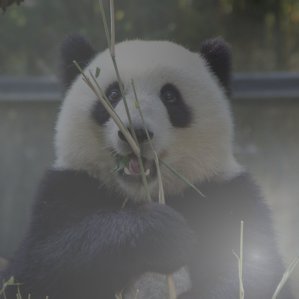}
    \caption{Fog}
    \end{subfigure}\hfill
    \centering
    \begin{subfigure}[b]{0.19\textwidth}
    \centering
    \includegraphics[width=\textwidth]{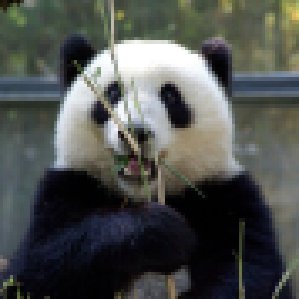}
    \caption{Pixelate}
    \end{subfigure}
    \caption{Visualization examples of commonly observable corruptions shown in Table~\ref{tab:corruptions}. \label{fig:corruption_via}}
\end{figure*}

\vspace{3pt}
\noindent\textbf{Corrupted data.}
The vulnerability to a small perturbation in inputs adversely affects the deployment of deep learning vision systems in many IoT applications that are sensitive to safety and user privacy. 
To quantitatively measure the robustness of the proposed \ourmethod{}, we consider the CIFAR-10-C dataset proposed by Hendrycks and Dietterich~\cite{hendrycks2018benchmarking}, who applied common observable corruption to the original (i.e., clean) test images of CIFAR-10. 
There are 19 different types of corruption from four main categories. 
See Table~\ref{tab:corruptions} for details and Fig.~\ref{fig:corruption_via} for visualization. 

Based on the empirical findings summarized in Table~\ref{tab:ood_robustness}, we observe that our \ourmethod{} performs significantly better than other peer models under a similar number of parameters. 
In addition, \ourmethod{} also performs noticeably better than the regular CNN model. 

\begin{table}[ht]
\centering
\caption{Robustness to commonly observable corruptions. We perform five runs and report mean performance along with standard deviation (mean\textcolor{darkgray}{$_{\pm\mbox{\footnotesize std}}$}). \label{tab:ood_robustness}}
\resizebox{.48\textwidth}{!}{%
    \begin{tabular}{@{\hspace{2mm}}l|cccc|c@{\hspace{2mm}}}
    \toprule
    Method & Noise & Blur & Weather & Digital & mean \\ \midrule
    CNN & 57.05\textcolor{darkgray}{$_{\pm7.55}$} & 74.90\textcolor{darkgray}{$_{\pm10.5}$} & 86.27\textcolor{darkgray}{$_{\pm5.41}$} & 82.32\textcolor{darkgray}{$_{\pm6.34}$} & 75.13\textcolor{darkgray}{$_{\pm12.9}$} \\
    LBCNN & 54.88\textcolor{darkgray}{$_{\pm6.87}$} & 67.61\textcolor{darkgray}{$_{\pm12.0}$} & 81.79\textcolor{darkgray}{$_{\pm7.49}$} & 78.41\textcolor{darkgray}{$_{\pm7.62}$} & 70.67\textcolor{darkgray}{$_{\pm12.1}$} \\
    PNN & 50.13\textcolor{darkgray}{$_{\pm6.67}$} & 63.27\textcolor{darkgray}{$_{\pm9.12}$} & 79.30\textcolor{darkgray}{$_{\pm7.56}$} & 76.09\textcolor{darkgray}{$_{\pm7.57}$} & 67.20\textcolor{darkgray}{$_{\pm11.5}$} \\
    GhostNet & 58.48\textcolor{darkgray}{$_{\pm5.93}$} & 67.53\textcolor{darkgray}{$_{\pm9.95}$} & 81.57\textcolor{darkgray}{$_{\pm7.35}$} & 78.10\textcolor{darkgray}{$_{\pm7.61}$} & 71.42\textcolor{darkgray}{$_{\pm9.08}$} \\ \midrule
    \ourmethod{} & 64.41\textcolor{darkgray}{$_{\pm5.42}$} & 77.41\textcolor{darkgray}{$_{\pm9.41}$} & 85.31\textcolor{darkgray}{$_{\pm5.01}$} & 82.22\textcolor{darkgray}{$_{\pm6.22}$} & \textbf{77.34\textcolor{darkgray}{$_{\pm7.98}$}} \\ \bottomrule
    \end{tabular}%
}
\end{table}

\begin{figure}[ht]
    \centering
    \includegraphics[width=.45\textwidth]{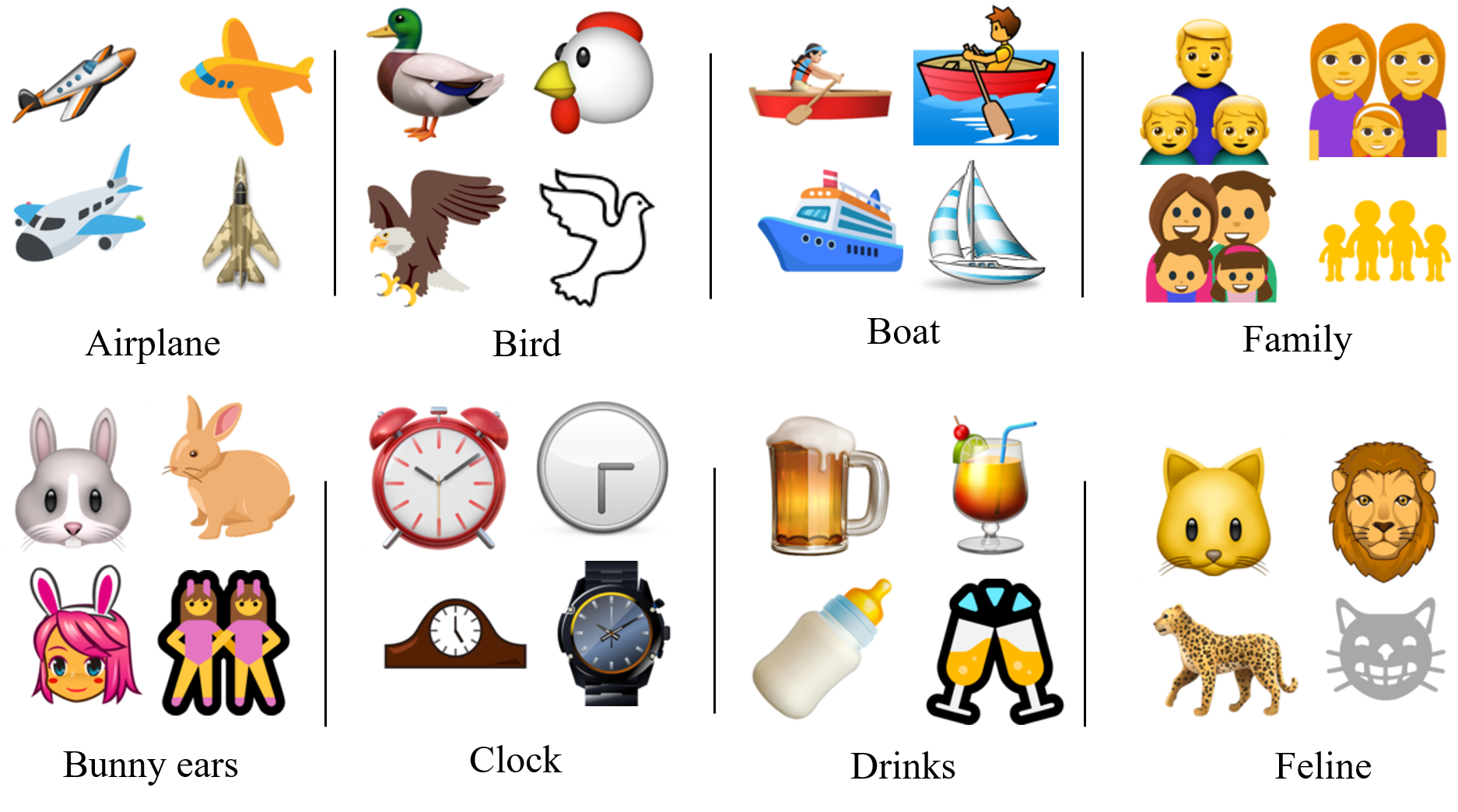}
    \caption{Visualization examples of the Icons-50 dataset. For each class, we show images collected from Apple, Facebook, Google, and Samsung from top-left to bottom-right. \label{fig:icons}}
\end{figure}

\vspace{3pt}
\noindent\textbf{Data under different styles.} In addition to data under degraded quality, another important angle for measuring robustness is the generalization performance on data under different styles, i.e., data with the same context but represented differently. We consider the Icons-50 dataset~\cite{hendrycks2018benchmarking}, which consists of 10K images from 50 classes of icons (e.g., airplane, symbols, activities, etc.) collected by various technology companies (e.g., Apple, Facebook, Google, etc.). We hold off data from one company while training on data from other companies to quantify robustness under different styles.
See Fig.~\ref{fig:icons} for a visualization.

As shown in Table~\ref{tab:style_robustness}, the mean accuracy of our proposed \ourmethod{} outperforms the mean accuracy of other models.
For example, our proposed \ourmethod{} achieves a mean accuracy improvement of 1.15\% compared to the regular CNN.
The main reason is that the parameters of \ourmethod{} are affected by both the filter generation function and the training data, which makes \ourmethod{} promising for achieving better performance than regular CNNs when dealing with test data whose style is inconsistent with the training data.

\begin{table}[ht]
\centering
\caption{Robustness to different styles. We perform five runs and report mean performance along standard deviation (mean\textcolor{darkgray}{$_{\pm\mbox{\footnotesize std}}$}).\label{tab:style_robustness}}
\resizebox{.48\textwidth}{!}{%
\begin{tabular}{@{\hspace{2mm}}l|cccc|c@{\hspace{2mm}}}
\toprule
Method & Apple & Facebook & Google & Samsung & Mean \\ \midrule
CNN & 91.74\textcolor{darkgray}{$_{\pm0.65}$} & 86.56\textcolor{darkgray}{$_{\pm0.25}$} & 82.63\textcolor{darkgray}{$_{\pm0.88}$} & 81.30\textcolor{darkgray}{$_{\pm1.11}$} & 85.56\textcolor{darkgray}{$_{\pm4.69}$} \\
LBCNN & 92.73\textcolor{darkgray}{$_{\pm0.67}$} & 87.63\textcolor{darkgray}{$_{\pm1.33}$} & 83.42\textcolor{darkgray}{$_{\pm0.40}$} & 79.09\textcolor{darkgray}{$_{\pm0.62}$} & 85.72\textcolor{darkgray}{$_{\pm5.83}$} \\
PNN & 92.49\textcolor{darkgray}{$_{\pm0.48}$} & 82.43\textcolor{darkgray}{$_{\pm1.37}$} & 82.24\textcolor{darkgray}{$_{\pm1.11}$} & 82.19\textcolor{darkgray}{$_{\pm1.67}$} & 84.84\textcolor{darkgray}{$_{\pm5.10}$} \\
GhostNet & 92.95\textcolor{darkgray}{$_{\pm0.80}$} & 85.26\textcolor{darkgray}{$_{\pm2.20}$} & 80.85\textcolor{darkgray}{$_{\pm0.58}$} & 76.64\textcolor{darkgray}{$_{\pm0.71}$} & 83.93\textcolor{darkgray}{$_{\pm6.97}$} \\ \midrule
MonoCNN & 93.52\textcolor{darkgray}{$_{\pm0.59}$} & 86.48\textcolor{darkgray}{$_{\pm0.66}$} & 82.42\textcolor{darkgray}{$_{\pm1.61}$} & 84.40\textcolor{darkgray}{$_{\pm0.87}$} & \textbf{86.71\textcolor{darkgray}{$_{\pm4.84}$}} \\ \bottomrule
\end{tabular}%
}
\end{table}

In Table~\ref{tab:ood_robustness} and Table~\ref{tab:style_robustness}, we observe that when there exists sufficient training data and the test data are within the same underlining distribution as the training data, all efficiency-oriented methods (i.e., LBCNN, PNN, GhostNet, and \ourmethod{}) exhibit a lower performance due to lower model capacity from limited parameters. However, the proposed FGF mechanism provides an inductive bias to the training of \ourmethod{}, which prevents overfitting to the training data, in turn, leading to a better generalization performance under limited training data and on out-of-distribution test data (i.e., corrupted data or data under different styles).

\subsection{Monomial Function Hyperparameter Study}

As demonstrated in the previous sections, we empirically observe that the monomial transformation is better suited for the filter generation function. 
In this section, we perform parameter sensitivity analysis on the hyperparameters of the monomial transformation. 

\begin{figure}[ht]
    \centering
    \includegraphics[width=.45\textwidth]{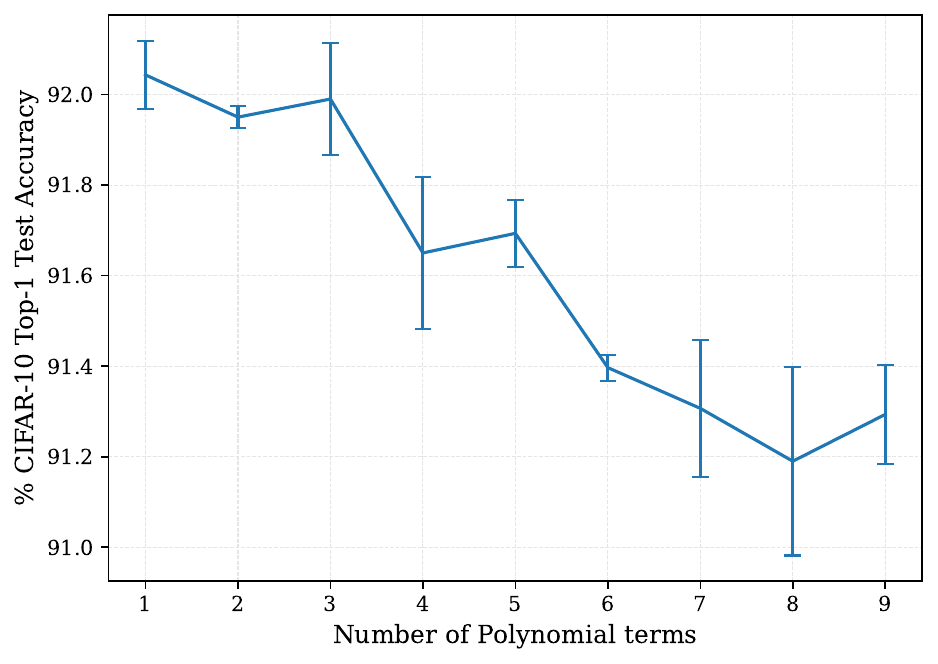}
    \caption{Impact of the number of terms used in polynomial transformation, where monomial transformation corresponds to the number of terms equal to one.\label{fig:abl_num_terms}}
\end{figure}

\vspace{3pt}
\noindent\textbf{Effect of polynomial terms.}
Instead of a monomial, one may relax the constraint on the number of terms to include the more general case of polynomial transformation. 
Accordingly, we allow the number of terms to grow from one (i.e., monomial) to many terms and evaluate the performance of corresponding models on CIFAR-10 classification. 
We repeat each setup five times and present the results in Fig.~\ref{fig:abl_num_terms}. 
We observe that monomial transformation (i.e., number of terms equal to one) is better suited for filter generation function as opposed to polynomial transformation with many terms.

\begin{figure}[ht]
    \centering
    \includegraphics[width=.49\textwidth]{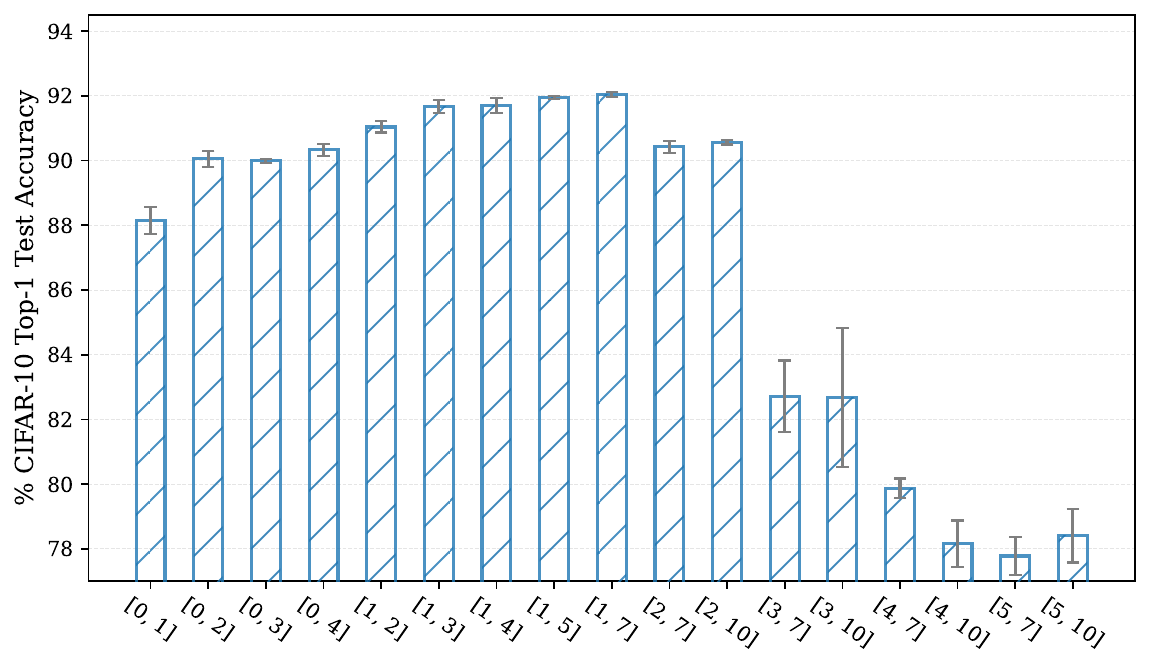}
    \caption{Impact of polynomial exponent range. The monomial exponent $\beta$ is uniformly sampled from $[a, b]$, where $a$ and $b$ are the lower and upper bounds.\label{fig:abl_exp_bound}}
    \vspace{-0.5cm}
\end{figure}

\vspace{3pt}
\noindent\textbf{Effect of monomial exponent.} Recall that we adopt the pointwise polynomial transformation as the filter generation function based on our empirical experiments. 
The monomial filter generation function randomly samples a (continuous-valued) exponent $\beta$ from $[a, b]$, where $a$ and $b$ are the lower and upper bounds on $\beta$. 
To understand the effect of $\beta$, we set the number of channels to 64 and the number of layers to 20 for our \ourmethod{} and vary the lower and upper bounds of $\beta$. 
Fig.~\ref{fig:abl_exp_bound} depicts the results. 
In general, having a diverse set of exponents $\beta$ (\ie, a larger range of $\beta$ bounds) leads to better performance of \ourmethod{}. 
Empirically, we identify that setting the lower bound $a$ to 1 and the upper bound $b$ to 7 yields the best performance.

%
\section{Conclusions and Future work} \label{ref:5-conclusion}
In this paper, we make the following two contributions.
First, we propose cloud-assisted training of a CNN model framework for IoT devices by considering model parameter transmission and model robustness.
Second, we propose a novel CNN architecture ({\emph{i.e.}}, \ourmethod{}) that reduces the number of model parameters sent by the cloud server to IoT devices by specifying only one filter that needs to be learned in each layer of \ourmethod{} and improves the robustness of the model by regularizing the model parameters using the filter generation function.
Experimental results show that the proposed approach achieves better performance in dealing with corrupted data and minimizes model parameter transmission.

In addition, Gill et al.~\cite{Gill@AI} comprehensively combed the emerging trends and future directions of AI for next-generation computing, which motivates our future work to start from the following points:
\begin{itemize}

\item We will deploy \ourmethod{} on the IoT device (such as the Raspberry Pi 4B) and test the resources and time it takes to generate \ourmethod{} based on learnable parameters, seeds, and filter generation function.

\item Since the available resources of the IoT device are dynamically changing, we need to deploy multiple \ourmethod{} variants with different capacities.
However, this faces two challenges: (i) how to divide multiple \ourmethod{} variants with different capacities and how to train these \ourmethod{} variants; (ii) how to reduce the storage resources occupied by deploying multiple \ourmethod{} variants.

\item IoT devices usually run multiple applications simultaneously. However, resources are limited. When the IoT device cannot provide sufficient resources for each application at the same time, how to reasonably allocate resources for each application poses a challenge.

\item Training the high-performance \ourmethod{} requires a large quantity of labeled data; however, unlabeled data are common in real scenarios, and how to train \ourmethod{} with the help of unlabeled data is a practical challenge.

\item To avoid leakage of user-sensitive private data, training \ourmethod{} on the IoT device is a research direction; however, how to speed up the training of \ourmethod{} is a challenge. 
\end{itemize}

\section{Acknowledgements}
This work was supported by the Fundamental Research Funds for the Central Universities (2021RC272), the National Natural Science Foundation of China (62106097), the China Postdoctoral Science Foundation (2021M691424, 2021M700364), the Research Grants Council of Hong Kong through the Theme-based Research Scheme (T-41-603/20R), and the Research Grants Council of Hong Kong through the General Research Fund (PolyU 15217919).

%


\ifCLASSOPTIONcaptionsoff
  \newpage
\fi

\bibliographystyle{IEEEtran}
\bibliography{IEEEexample}

\end{document}